\documentclass[11pt]{article}

\usepackage[T1,T2A]{fontenc}
\usepackage[cp1251]{inputenc}
\usepackage{textcomp}
\usepackage[russian,english]{babel}
\usepackage[centertags]{amsmath}
\usepackage[mediummath]{nccmath}
\usepackage{amsfonts}
\usepackage{amssymb}
\usepackage[pdftex]{hyperref}
\usepackage{graphicx}
\usepackage[numbers,sort&compress]{natbib}

\usepackage{paperinitial}

\paperinitialization{15mm}{15mm}{15mm}{15mm}{2pt}{10pt}

\DeclareMathOperator{\re}{Re}
\DeclareMathOperator{\im}{Im}
\DeclareMathOperator{\sgn}{sgn}

\DeclareMathOperator{\rot}{rot}
\DeclareMathOperator{\diag}{diag}

\newcommand{\bs}{\boldsymbol}

\newcommand{\e}{\varepsilon}
\newcommand{\vf}{\varphi}
\newcommand{\vk}{\varkappa}
\newcommand{\s}{\sigma}

\newcommand{\al}{\alpha}
\newcommand{\be}{\beta}
\newcommand{\ga}{\gamma}

\newcommand{\de}{\delta}
\newcommand{\De}{\Delta}

\newcommand{\la}{\lambda}

\newcommand{\ups}{\upsilon}

\newcommand{\spx}{\mathbf{x}}

\newcommand{\spk}{\mathbf{k}}
\newcommand{\spe}{\mathbf{e}}

\begin{document}\selectlanguage{english}
\allowdisplaybreaks[4]
\frenchspacing
\setlength{\unitlength}{1pt}

\title{{\Large\textbf{Short wavelength band structure of photons in cholesteric liquid crystals}}}

\date{}

\author{
O.V. Bogdanov${}^{1),2)}$\thanks{E-mail: \texttt{bov@tpu.ru}},\;
P.O. Kazinski${}^{1)}$\thanks{E-mail: \texttt{kpo@phys.tsu.ru}},\;
P.S. Korolev${}^{1)}$\thanks{E-mail: \texttt{kizorph.d@gmail.com}},\;
and G.Yu. Lazarenko${}^{1)}$\thanks{E-mail: \texttt{laz@phys.tsu.ru}}\\[0.5em]
{\normalsize ${}^{1)}$ Physics Faculty, Tomsk State University, Tomsk, 634050, Russia}\\
{\normalsize ${}^{2)}$ Tomsk Polytechnic University, Tomsk, 634050, Russia}
}

\maketitle

\begin{abstract}

The band structure of photons in cholesteric liquid crystals (CLCs) is investigated in the shortwave approximation. The bound states or narrow resonances of photons in the CLC are formed by the extraordinary waves. The explicit expressions for the spectrum bands and the dispersion laws of photons in these bands are obtained. It is shown that these states describe photons propagating almost perpendicular to the CLC axis. The density of photon states acquires a sharp peak due to the presence of bound states. Near this peak, in the particular case of plasma permittivity, the photons posses a linear or quadratic dispersion relations in the directions perpendicular to the CLC axis depending on the sign of the anisotropy of the CLC permittivity tensor. The resonances in the CLC plate are also described.

\end{abstract}

\section{Introduction}

The optical properties of the cholesteric liquid crystals (CLCs) are a well studied subject \cite{BelDmitOrl,BelSon82,deGennProst,Andri,BelyakovBook,VetTimShab20}. However, owning to the fact that the Maxwell equation are not exactly solvable for the CLC permittivity tensor for arbitrary angles of photon propagation, the investigation of its band structure encounters with significant difficulties. It is well known \cite{BelDmitOrl,BelSon82,deGennProst,Andri,BelyakovBook,VetTimShab20,Kats,Mauguin11,deVries51} that, when the propagation angle to the CLC axis is small, the spectrum of photons in CLCs possesses two allowed bands and a singe forbidden band. As the propagation angle to the CLC axis increases, there appear additional forbidden bands in the spectrum \cite{BelDmitOrl,BelSon82,BelyakovBook,BerrSchef,Takezoe83,TakezoeH83,OldMirVal84,RisSchm19}. This band structure is described numerically in a certain range of parameters \cite{BerrSchef,Takezoe83,TakezoeH83,OldMirVal84,RisSchm19} and analytically in the framework of perturbation theory with respect to the anisotropy of the CLC permittivity tensor and in the two-wave approximation \cite{BelDmitOrl,BelSon82,BelyakovBook}. For sufficiently large photon energies, the perturbation theory with respect to anisotropy is poorly applicable\footnote{The breakdown of the perturbation theory with respect to anisotropy for large photon energies can easily be understood if one bears in mind the analogy of the permittivity tensor with the metric tensor in a curved space. In these terms, the cause of failure of the perturbation theory with respect to anisotropy is the same as the cause of inapplicability of the standard perturbation theory in quantum gravity for energy scales above the Planck scale.} (see for details \cite{parax,wkb_chol}) and so one needs to resort to other approximations. In the parameter domain where the energies of photons are large, the shortwave approximation suggests itself.

The shortwave approximation to the solutions of the Maxwell equations in CLCs was also studied \cite{Osadch84,Osadch85,AksValRom08,AksValRom01,AksValRom04,AksKryuRom06}. It was found in \cite{AksValRom08,AksValRom01,AksValRom04,AksKryuRom06} that there is a confinement of light modes to the directions perpendicular to the CLC axis. This theoretical prediction was confirmed experimentally in \cite{Aksenova05}. Nevertheless, to our knowledge, the explicit expressions for the spectrum of the bound states and resonances, the band structure, the dispersion laws, and the corresponding contributions to the density of states in the shortwave approximation have been unknown up to now. The aim of the present paper is to fill this gap. Of course, all these features can be described numerically. However, the dependence of the characteristics of the photon spectrum in CLCs on the relevant parameters is obscure in numerical investigations and so the explicit analytical expressions are needed. The aforementioned bound states resemble the bound states in thin-film waveguides. So, CLCs provide a natural realization of such waveguides or, more precisely, of a stack of such waveguides. Note however that the modes we are about to describe are distributed over the entire volume of the CLC. The surface modes in CLCs were described analytically in \cite{BelOrl91,Shiyan90}.

The bound states and resonances result in a rapid increase of the density of states that in turn leads to a growth of probability of quantum processes with photons in CLCs \cite{Sakoda}. In particular, we shall show that the resonances are responsible for the wavy structure in the spectrum of transition radiation produced by charged particles traversing a CLC plate  \cite{parax,wkb_chol}. It is clear that such an increase should be observed for other radiation processes. The theory that will be expounded in the present paper relies only on the peculiar form of the permittivity tensor of a CLC and therefore it is applicable to any material possessing the permittivity tensor of such a form in the corresponding parameter domain, for example, to specially designed metamaterials \cite{Sakoda,JJWMbook}.

In Sec. \ref{Max_Eqs}, we start with the description of the Maxwell equations in CLCs paying a special attention to the ultraviolet behaviour of the permittivity tensor. In Sec. \ref{WKB_Sols}, we briefly describe the solutions of the Maxwell equations in the CLC plate in the shortwave approximation and construct the quantum electromagnetic field in such a medium. The CLC plate is assumed to be orthogonal to the CLC axis with the width equal to an integer number of the CLC periods. The explicit expression for the quantum electromagnetic field allows one to investigate various quantum processes in this medium as, for example, the creation of photons by particles moving in it (see, e.g., \cite{BKL5,KazLaz20,parax,wkb_chol,GinzbThPhAstr,AbrGorDzyal}). Sections \ref{Bound_Stat} and \ref{Reson_Sec} constitute the main part of the paper. In Sec. \ref{Bound_Spec}, we derive the short wavelength band spectrum of photons in the CLC and, in Sec. \ref{Plasm_Perm}, we particularize the general formulas to the case of the permittivity tensor of a plasma-like form which is realized for high energy photons. In this case, we obtain the explicit expressions for the dispersion relations for the photons in the bands and find the corresponding contributions to the density of photon states. In Sec. \ref{Reson_Sec}, we analyze the resonances in the CLC plate and discuss their contribution to the intensity of transition radiation created by a charged particle crossing this plate. In Appendix \ref{Unit_Rels}, we collect the formulas for the scattering problem on the line for the matrix Schr\"{o}dinger equation. Appendix \ref{Join_Coeff_App} contains the explicit expressions for the joining coefficients of the mode functions in the CLC plate.

Throughout the paper we use the system of units such that $\hbar=c=1$ and $e^2=4\pi\al$, where $\al$ is the fine structure constant. We also denote the axes $x$, $y$, and $z$ as the axes $1$, $2$, and $3$, respectively, and $x^0\equiv t$ is the time variable.

\section{Maxwell equations}\label{Max_Eqs}

The simplest model of a nematic in the cholesteric phase is described by the permittivity tensor of the form \cite{BelDmitOrl,BelSon82,deGennProst,Andri,BelyakovBook,VetTimShab20}
\begin{equation}\label{permit_holec}
    \e_{ij}(k_0)=\e_\perp(k_0)\de_{ij}+(\e_\parallel(k_0)-\e_\perp(k_0))\tau_i\tau_j.
\end{equation}
The director changes periodically as
\begin{equation}
    \tau_i=(\cos(qz),\sin(qz),0),
\end{equation}
where $q$ characterizes the period of director variations. Below, for definiteness, we suppose that $q>0$. The quantities $\e_{\parallel,\perp}(k_0)$ were studied experimentally for prevalent CLCs in the optical range and below on the photon energy scale \cite{LiWu,LWGLW,OldMirVal84,TakezoeH83,Takezoe83}. For high energy photons, we suppose that
\begin{equation}\label{permit_UV}
    \e_{\perp}(k_0)=1-\omega_{\perp}^2/k_0^2,\qquad \e_{\parallel}(k_0)=1-\omega_{\parallel}^2/k_0^2.
\end{equation}
The more accurate formula relating $\e_{\parallel,\perp}(k_0)$ to molecular polarizabilities is given in \cite{LiWu,LWGLW,Vuks66}. When $\e_{\parallel,\perp}\approx1$, this formula turns into \eqref{permit_UV} for photons with sufficiently large energies. In order to estimate the plasma frequency, $\omega_{\parallel,\perp}\sim\omega_p$, one can employ the formula
\begin{equation}
    \omega_p^2=4\pi\al Z_m n_m/m_e,\qquad n_m=\rho_m/(m_p M_r),
\end{equation}
where $Z_m$ is the number of electrons in the molecule, $M_r$ is the relative molecular mass, $\rho_m$ is the mass density, $n_m$ is the concentration of molecules, $m_e$ is the electron mass, and $m_p$ is the proton mass. For example, for the mass density $1$ g/cm${}^3$, the plasma frequency of the cholesteric C${}_{18}$H${}_{19}$N is equal to $21$ eV. Furthermore, one should keep in mind that in liquid crystals consisting of rod-shaped molecules the director $\tau_i$ is directed along the long axis of the molecules arranged in the helical structure. As a rule, the molecular polarizability is larger along the long axis of the molecule than along its short transverse axes \cite{YangWu06}. Thus, for such molecules and CLCs we have $\omega_\parallel>\omega_\perp$. As for the CLCs consisting of disc-shaped molecules, $\omega_\parallel<\omega_\perp$. The typical sizes of a cholesteric molecule are $2\times0.5$ nm. Hence, the permittivity tensor \eqref{permit_holec} can be used to describe the electromagnetic properties of CLCs for the photon energies $k_0\lesssim100$ eV.

The quantity
\begin{equation}\label{anisotropy}
    \de\e(k_0):=(\e_\parallel(k_0)-\e_\perp(k_0))/\e_\perp(k_0)
\end{equation}
characterizes the anisotropy of the CLC permittivity tensor. As a rule, it is small and can take positive or negative values. We suppose that the CLC constitutes a plate placed in a vacuum. This plate is perpendicular to the $z$ axis and has the width $L=\pi N/q$, where $N$ is the number of periods. It is the form of the dielectric tensor \eqref{permit_holec} which is only relevant for our further considerations. Such a permittivity tensor can describe the electromagnetic properties of the (meta)materials different from CLCs and all the conclusions that we will draw regard those materials as well.

In order to describe the quantum dynamics of excitations in the plate of CLC, we construct the quantum electromagnetic field in this medium. To this end, in accordance with the general procedure expounded, for example, in \cite{BKL5,parax,GinzbThPhAstr,AbrGorDzyal}, one needs to find a complete set of solutions of the Maxwell equations
\begin{equation}\label{Max_eqns}
    \big[k_0^2\e_{ij}(k_0,\spx)-\rot^2_{ij}\big]A_j(k_0,\spx)=0,\qquad\partial_i(\e_{ij}(k_0,\spx)A_j(k_0,\spx))=0.
\end{equation}
The second equation is the generalized Coulomb gauge. It follows from the first equation in \eqref{Max_eqns} for $k_0\neq0$. Henceforth, we suppose that the electromagnetic field obeys the boundary conditions that exclude the zero mode, viz., the first equation in \eqref{Max_eqns} does not have solutions for $k_0=0$.

\section{Short wavelength solutions}\label{WKB_Sols}

In the general case, i.e., for arbitrary values of quantum numbers of the complete set, it seems impossible to construct the exact solutions to the Maxwell equations \eqref{Max_eqns} in a CLC. In the paper \cite{wkb_chol}, we found the complete set of solutions of \eqref{Max_eqns} in the CLC plate in the shortwave approximation (see also \cite{AksValRom08,AksValRom01,AksValRom04,AksKryuRom06}). This region of parameters is usually called the Mauguin limit \cite{Mauguin11}. For the reader convenience we provide here the explicit expressions for the semiclassical mode functions in the leading order in $1/k_0$. The general theory of the matrix shortwave approximation can be found in \cite{BabBuld,Maslov,BagTrif,RMKD18}.

To this end, we introduce the basis $\{\spe_+,\spe_-,\spe_3\}$ with
\begin{equation}
    \spe_\pm:=\spe_1\pm i \spe_2.
\end{equation}
In this basis, any vector $\mathbf{A}$ has the form
\begin{equation}\label{cylin_basis}
    \mathbf{A}=\frac{1}{2}(\spe_+ A_-+\spe_- A_+)+\spe_3 A_3, \qquad A_3=(\spe_3,\mathbf{A}),\quad A_\pm=(\spe_\pm,\mathbf{A}).
\end{equation}
The translation invariance of \eqref{permit_holec} along the axes $x$ and $y$ allows one to write the complete set of solutions of \eqref{Max_eqns} as
\begin{equation}\label{pln_waves}
    A_i(k_0,\spx)=e^{i\spk_\perp\spx_\perp}A_i(k_0,z).
\end{equation}
Introducing the notation
\begin{equation}\label{apm_not}
    a_\pm=A_\pm e^{\mp i\vf},\qquad A_3\equiv a_3,
\end{equation}
for the components of $A_i(k_0,z)$, where $\vf:=\arg k_+$, the Maxwell equations are reduced to the matrix Sch\"{o}dinger equation (see for details \cite{parax})
\begin{equation}\label{Max_eqns2}
    (\partial_3K\partial_3+V)
        \left[
      \begin{array}{c}
        a_+ \\
        a_- \\
      \end{array}
    \right]=0,
\end{equation}
where
\begin{equation}
    K=\left[
      \begin{array}{cc}
         1+\frac{k_\perp^2}{2\bar{k}_3^2} & \frac{k_\perp^2}{2\bar{k}_3^2} \\
         \frac{k_\perp^2}{2\bar{k}_3^2} & 1+\frac{k_\perp^2}{2\bar{k}_3^2} \\
      \end{array}
    \right],\qquad
    V=-
    \frac{k_\perp^2}{2}
    \left[
      \begin{array}{cc}
         1 & -1 \\
         -1 & 1 \\
      \end{array}
    \right]+
    \frac{\bar{k}_0^2}{2}
    \left[
      \begin{array}{cc}
        2+\de\e & \de\e e^{2i\bar{\theta}} \\
        \de\e e^{-2i\bar{\theta}} & 2+\de\e \\
      \end{array}
    \right],
\end{equation}
and $\bar{k}_3^2:=\bar{k}_0^2-k_\perp^2$, $\bar{k}_0^2:=k_0^2\e_\perp$. The quantity $\bar\theta:=qz-\vf$ is the angle between the director $\bs\tau$ and the projection of the photon momentum $\spk_\perp$. The rest component of the potential is given by
\begin{equation}\label{a3_not}
    a_3=\frac{ik_\perp}{2(\bar{k}_0^2-k_\perp^2)}\partial_3(a_++a_-).
\end{equation}
Solving Eq. \eqref{Max_eqns2} in the shortwave approximation \cite{AksValRom08,AksValRom01,AksValRom04,AksKryuRom06,wkb_chol}, we obtain the ordinary wave
\begin{equation}\label{ordinary_+}
    k_3^{(1)}=\bar{k}_3,\qquad
    \left[
      \begin{array}{c}
        a^{(1)}_+ \\
        a^{(1)}_- \\
      \end{array}
    \right]=
    \Big(\frac{\bar{k}_3}{2(\bar{k}_0^2-k_\perp^2\cos^2\bar{\theta})}\Big)^{1/2}
    \left[
      \begin{array}{c}
        ie^{i\bar{\theta}} \\
        -ie^{-i\bar{\theta}} \\
      \end{array}
    \right]e^{i\bar{k}_3z},\qquad a_3^{(1)}=\frac{k_\perp \sin\bar{\theta} e^{i\bar{k}_3z}}{\sqrt{2\bar{k}_3(\bar{k}_0^2-k_\perp^2\cos^2\bar{\theta})}},
\end{equation}
the reflected ordinary wave
\begin{equation}\label{ordinary_-}
    \left[
      \begin{array}{c}
        b^{(1)}_+ \\
        b^{(1)}_- \\
      \end{array}
    \right]=
    \Big(\frac{\bar{k}_3}{2(\bar{k}_0^2-k_\perp^2\cos^2\bar{\theta})}\Big)^{1/2}
    \left[
      \begin{array}{c}
        ie^{i\bar{\theta}} \\
        -ie^{-i\bar{\theta}} \\
      \end{array}
    \right]e^{-i\bar{k}_3z},\qquad b_3^{(1)}=\frac{-k_\perp \sin\bar{\theta} e^{-i\bar{k}_3z}}{\sqrt{2\bar{k}_3(\bar{k}_0^2-k_\perp^2\cos^2\bar{\theta})}},
\end{equation}
the extraordinary wave
\begin{equation}\label{extraordinary_+}
\begin{gathered}
    k_3^{(2)}=\sqrt{(1+\de\e)\bar{k}_3^2+\de\e k_\perp^2\sin^2\bar{\theta}},\qquad
    \left[
      \begin{array}{c}
        a^{(2)}_+ \\
        a^{(2)}_- \\
      \end{array}
    \right]=
    \Big(2k_3^{(2)}\bar{k}_0^2(\bar{k}_0^2-k_\perp^2\cos^2\bar{\theta})\Big)^{-1/2}
    \left[
      \begin{array}{c}
        \zeta^* \\
        \zeta \\
      \end{array}
    \right]e^{iS(\bar{\theta})},\\
    a_3^{(2)}=-k_\perp\cos\bar{\theta}\Big(\frac{k_3^{(2)}}{2\bar{k}_0^2(\bar{k}_0^2-k_\perp^2\cos^2\bar{\theta})}\Big)^{1/2} e^{iS(\bar{\theta})},
\end{gathered}
\end{equation}
where
\begin{equation}
\begin{gathered}
    \zeta=\bar{k}_3^2\cos\bar{\theta} -i\bar{k}_0^2\sin\bar{\theta},\qquad S=q^{-1}\sqrt{(1+\de\e)\bar{k}_3^2}E\Big(\bar{\theta};\frac{-\de\e k_\perp^2}{(1+\de\e)\bar{k}_3^2}\Big),\\
    E(\tau;\vk):=\int_0^\tau d \xi\sqrt{1-\vk\sin^2\xi},
\end{gathered}
\end{equation}
and the reflected extraordinary wave
\begin{equation}\label{extraordinary_-}
\begin{split}
    \left[
      \begin{array}{c}
        b^{(2)}_+ \\
        b^{(2)}_- \\
      \end{array}
    \right]&=
    \Big(2k_3^{(2)}\bar{k}_0^2(\bar{k}_0^2-k_\perp^2\cos^2\bar{\theta})\Big)^{-1/2}
    \left[
      \begin{array}{c}
        \zeta^* \\
        \zeta \\
      \end{array}
    \right]e^{-iS(\bar{\theta})},\\
    b_3^{(2)}&=k_\perp\cos\bar{\theta}\Big(\frac{k_3^{(2)}}{2\bar{k}_0^2(\bar{k}_0^2-k_\perp^2\cos^2\bar{\theta})}\Big)^{1/2} e^{-iS(\bar{\theta})}.
\end{split}
\end{equation}
In evaluating $a_3$ by means of \eqref{a3_not}, only the leading in $1/k_0$ term is taken into account. The shortwave approximation used is applicable when
\begin{equation}\label{short_wave2}
    k_0|\de\e|\gg q.
\end{equation}
In particular, the above semiclassical mode functions are not valid in the limit $\de\e\rightarrow0$. It is also clear that these solutions do not reproduce the well-known forbidden energy band when the photon propagation angle to the CLC axis is small.

The electromagnetic waves with $n_\perp:=k_\perp/k_0<1$ pass through the CLC plate. These modes should be joined with the solutions of the Maxwell equations out of the plate using the standard boundary conditions
\begin{equation}\label{bound_conds_chol}
    [A_\pm]_{z=0}=[A_\pm]_{z=-L}=0,\qquad [\rot A_\pm]_{z=0}=[\rot A_\pm]_{z=-L}=0,
\end{equation}
where
\begin{equation}
    \rot A_\pm=\pm i(\partial_3A_\pm -\partial_\pm A_3).
\end{equation}
In evaluating the curl of the semiclassical modes \eqref{ordinary_+}-\eqref{extraordinary_-}, it is necessary to take into account only such terms that are leading in powers of $1/k_0$, These terms arise when the derivative acts on the rapidly oscillating exponent.

To distinguish the operator of the quantum electromagnetic field $\hat{\mathbf{A}}(x)$ from the mode functions, we will denote the latter as $\bs{\psi}(\spx)$. Then, for $z>0$, the mode function $\bs{\psi}(s,\spk;x)$ of the quantum electromagnetic field has the form
\begin{equation}\label{mode_funcz>0}
    \bs{\psi}(s,\spk;\spx)=\frac{c}{\sqrt{2k_0V}}\mathbf{f}(s,\spk)e^{-ik_0x^0+i\spk\spx},
\end{equation}
where the polarization vector of a photon with helicity $s$ in the Cartesian basis is written as
\begin{equation}
    \mathbf{f}(s,\spk)=(n_3 \cos\vf -is\sin\vf,n_3\sin\vf +is\cos\vf,-n_\perp)/\sqrt{2},
\end{equation}
and $n_3=\cos\theta$. In the CLC plate, the mode function is given by
\begin{equation}\label{mode_func_cho}
    \bs{\psi}(s,\spk;\spx)=\frac{c}{\sqrt{2k_0V}}[r_1\bs{\psi}^{(1)}_r+r_2\bs{\psi}^{(2)}_r+l_1\bs{\psi}^{(1)}_l+l_2\bs{\psi}^{(2)}_l]e^{-ik_0 x^0+i\spk_\perp \spx_\perp},
\end{equation}
where the components $\bs{\psi}^{(1)}_r$, $\bs{\psi}^{(2)}_r$, $\bs{\psi}^{(1)}_l$, and $\bs{\psi}^{(2)}_l$ in the basis \eqref{cylin_basis} take the form \eqref{ordinary_+}, \eqref{extraordinary_+}, \eqref{ordinary_-}, and \eqref{extraordinary_-}, respectively, and one should take into account the notation \eqref{apm_not}. For $z<-L$, the mode function is
\begin{equation}\label{mode_funcz<mL}
    \bs{\psi}(s,\spk;\spx)=\frac{c}{\sqrt{2k_0V}}\big\{[d_+\mathbf{f}_{++} +d_-\mathbf{f}_{-+}]e^{ik_3z} +[h_+\mathbf{f}_{+-} +h_-\mathbf{f}_{--}]e^{-ik_3z}\big\}e^{-ik_0x^0+i\spk_\perp\spx_\perp},
\end{equation}
where
\begin{equation}
    \mathbf{f}_{++}:=\mathbf{f}(1,\spk),\qquad \mathbf{f}_{-+}:=\mathbf{f}(-1,\spk),\qquad \mathbf{f}_{+-}:=\mathbf{f}(1,\spk_\perp,-k_3),\qquad \mathbf{f}_{--}:=\mathbf{f}(-1,\spk_\perp,-k_3).
\end{equation}
The constant $c$ is found from the normalization condition. In fact, the solution \eqref{mode_funcz>0}, \eqref{mode_func_cho}, \eqref{mode_funcz<mL} is the shortwave approximation to the Jost function $F^+(s,k_3)$ defined in \eqref{Jost_funcs}. The coefficients $r_i$, $l_i$, $d_\pm$, $h_\pm$ of the linear combinations and the explicit expression for the normalization constant $c$ are presented in Appendix \ref{Join_Coeff_App}. If $d_\pm$, $h_\pm$ are known in the form of analytic functions of $k_3$, then one can readily find the bound states and resonances of photons in the CLC plate (see Appendix \ref{Unit_Rels}).

It is supposed in the expressions for the extraordinary wave \eqref{extraordinary_+} that $k_3^{(2)}$ is nonnegative for any $\bar{\theta}$, i.e., the turning points are absent. The presence of turning points signifies that there are bound states of a photon in the CLC plate of an infinite width that are caused by total internal reflection. In the next section, we shall consider such states in detail where it will be shown that they appear when $n_\perp$ belongs to a certain interval (see inequalities \eqref{np_dem} and \eqref{np_dep}). For the range of values of $n_\perp$ where the turning point are absent, it is useful to rewrite the Hamilton-Jacobi action for the extraordinary wave as
\begin{equation}
    S(\bar{\theta})=:\bar{S}(\bar{\theta})+p_3\bar{\theta}/q,
\end{equation}
where
\begin{equation}\label{quasimom_p3}
    p_3=\frac{2}{\pi}E\Big(\frac{-\de\e k_\perp^2}{(1+\de\e)\bar{k}_3^2}\Big)\sqrt{1+\de\e}\,\bar{k}_3,\qquad E(\vk):=E(\pi/2;\vk).
\end{equation}
The function $\bar{S}(\bar{\theta})$ is a periodic function of $\bar{\theta}$ with the period $\pi$ and $p_3$ specifies the quasimomentum of the extraordinary wave.

The operator of the quantum electromagnetic field in the interaction representation reads
\begin{equation}\label{quantum_field}
    \hat{\mathbf{A}}(x)=\sum_{s=\pm1}\int\frac{V d\spk}{(2\pi)^3}\bs{\psi}(s,\spk;x)\hat{a}(s,\spk)+\sum_{\al} \bs{\psi}_\al(x)\hat{a}_\al+\text{H.c.},
\end{equation}
where $\hat{a}$, $\hat{a}^\dag$ are the creation-annihilation operators
\begin{equation}
    [\hat{a}(s,\spk),\hat{a}^\dag(s',\spk')]=\frac{(2\pi)^3}{V}\de_{s,s'}\de(\spk-\spk'),\qquad [\hat{a}_\al,\hat{a}^\dag_\be]=\de_{\al\be},
\end{equation}
and the second term on the right-hand side of \eqref{quantum_field} is responsible for the contribution of the bound states. These states possess $n_\perp>1$.

\section{Bound states}\label{Bound_Stat}
\subsection{Band spectrum}\label{Bound_Spec}

The equation \eqref{Max_eqns2} possesses the bound states in the CLC of an infinite width. In this section, we shall find the explicit expressions for these states in the shortwave approximation, their spectrum, and its splitting due to tunneling through a potential barrier.

To analyze the bound states, we have to consider the two cases $\de\e<0$ and $\de\e>0$ separately. We begin with the case $-1<\de\e<0$. However, many of the below formulas will be written in such a form that is valid for both cases. The Hamilton-Jacobi action is given by
\begin{equation}
    S(\bar{\theta})=\frac{1}{q}\int_0^{\bar{\theta}}d\tau k_3^{(2)}(\tau)=\frac{k_0}{q}\int_0^{\bar{\theta}} d\tau\sqrt{(1+\de\e)\bar{n}_3^2+\de\e n_\perp^2\sin^2\tau}=:k_0\s(\bar{\theta}),
\end{equation}
where $\bar{n}_3^2:=\e_\perp-n_\perp^2$. The turning point,
\begin{equation}\label{turn_point}
    \bar{\theta}_0=\arcsin\frac{1}{\sqrt{\vk}},\qquad \vk:=-\frac{\de\e n_\perp^2}{(1+\de\e)(\e_\perp-n_\perp^2)}=\frac{(\e_\perp-\e_\parallel) n_\perp^2}{\e_\parallel(\e_\perp-n_\perp^2)},
\end{equation}
appears when
\begin{equation}\label{np_dem}
    \e_\parallel\leqslant n_\perp^2\leqslant\e_\perp.
\end{equation}

The modes with $n_\perp<1$ tunnel from the CLC plate of a finite width to the vacuum where they propagate freely. The reverse process is also possible, i.e., the incident electromagnetic wave with $n_\perp$ from the interval \eqref{np_dem} and the given energy $k_0$ that will be found below tunnels through the CLC plate. It is clear from \eqref{np_dem} that this situation is possible only when $\e_\parallel<1$. Below, we will also refer to the modes with $n_\perp<1$ possessing the turning points as the bound states despite the fact that they have a finite lifetime for the CLC plate of a finite width, i.e., they are resonances.

\begin{figure}[t]
   \centering
   \includegraphics*[width=0.49\linewidth]{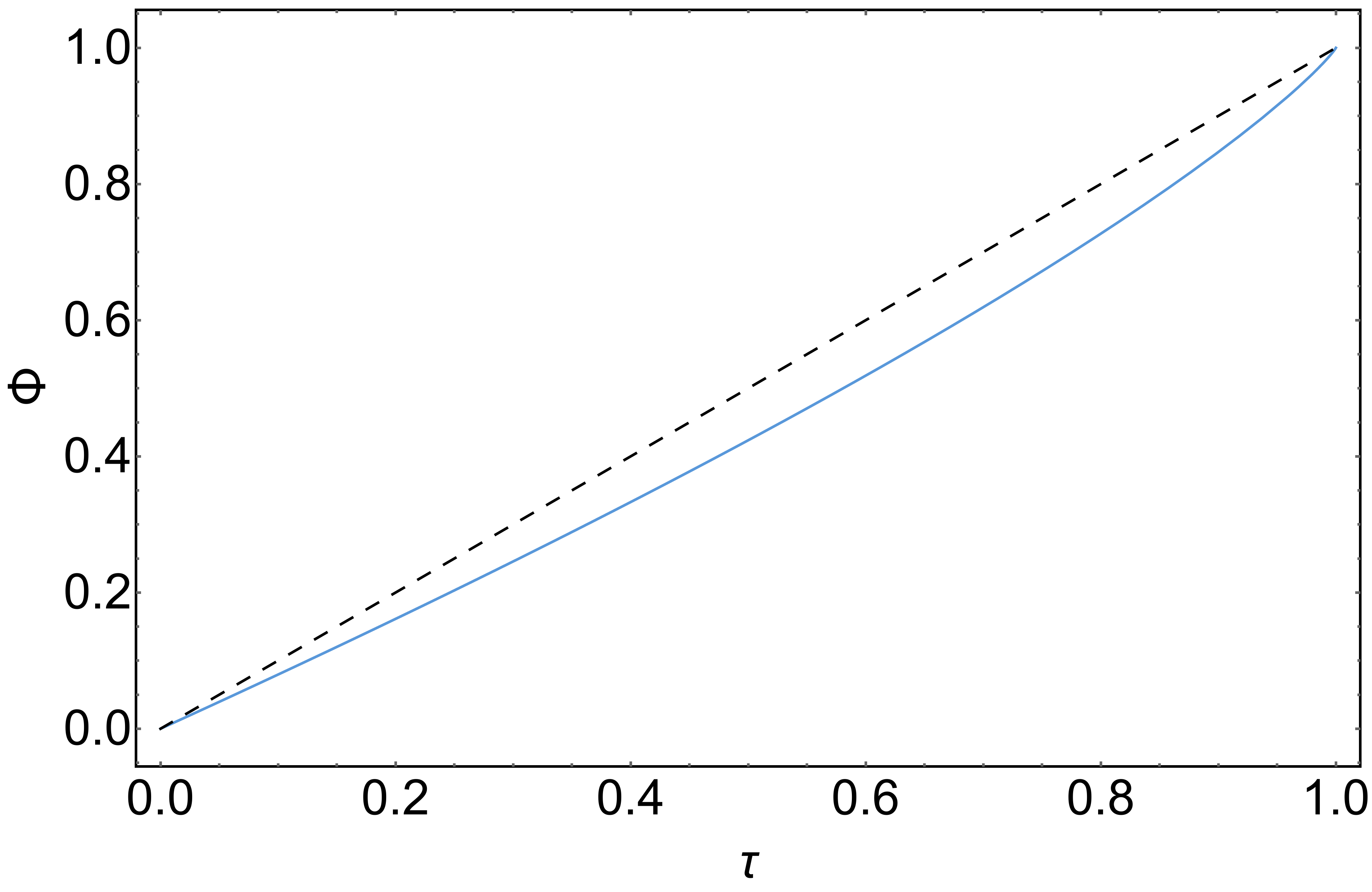}\;
   \includegraphics*[width=0.49\linewidth]{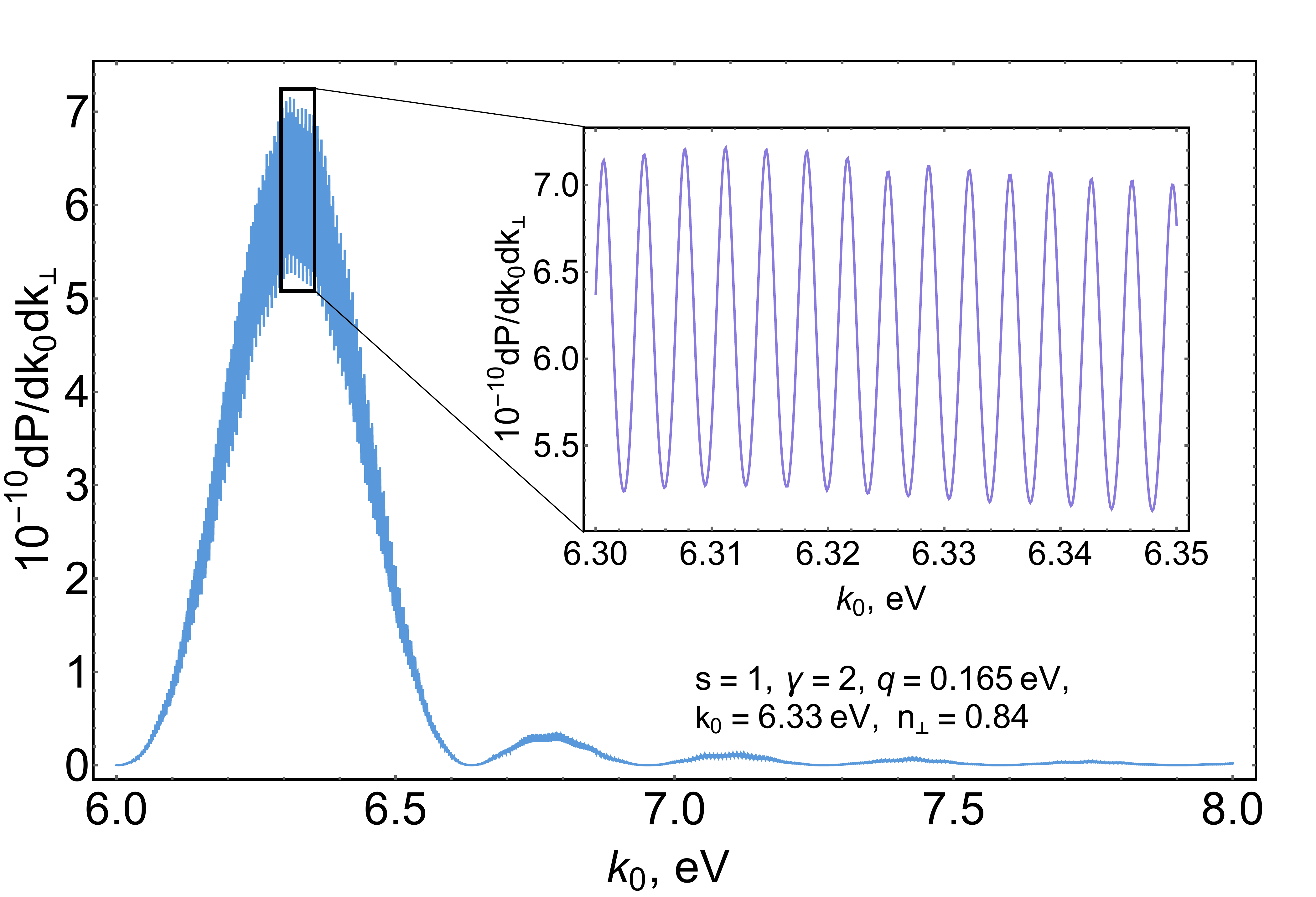}\;
   \caption{{\footnotesize On the left panel: The comparison of the functions $\Phi(\tau)$ (the solid line) and $\tau$ (the dashed line) for $\tau\in[0,1]$. On the right panel: The average number of plane-wave photons produced in transition radiation by a charged particle traversing normally the CLC plate. This is the enlarged upper left plot in Fig. 2 of \cite{wkb_chol}.}}
\label{Phi_dP_plots}
\end{figure}

The modes with $n_\perp>1$ cannot leave the CLC plate with an infinite frontal area because of total internal reflection. Nevertheless, the modes with $n_\perp>1$ can be observed in the vacuum if one changes the geometry of the CLC plate by tilting, for example, one of the facets of the plate to avoid total internal reflection, or these modes can be observed as escaping through the lateral faces of the actual CLC plate. In fact, these modes can be excited and be brought out as the bound modes in thin-film waveguides \cite{ZolKisSych}. Moreover, these modes and only them are excited by a charged particle moving parallel to the CLC plate at a small distance from it. The one-particle transition amplitude to the state with such a photon is proportional to
\begin{equation}
    \de(k_0-\spk_\perp \bs{\beta}_\perp)e^{-k_0d\sqrt{n_\perp^2-1}},
\end{equation}
where $d$ is the distance from the particle trajectory to the surface of the plate. The argument of the delta function vanishes when
\begin{equation}
    n_\perp=1/(\be\cos\theta),
\end{equation}
where $\theta$ is the angle between the vectors $\mathbf{n}_\perp$ and $\bs{\beta}$. In the ultrarelativistic limit, $\ga\gg1$, we obtain
\begin{equation}
    \sqrt{n_\perp^2-1}\approx\frac{\sqrt{1+\theta^2\ga^2}}{\ga},
\end{equation}
where $\theta\ll1$. Hence, the ultrarelativistic charged particle excites mostly the modes with $\mathbf{n}_\perp$ codirectional with the velocity $\bs{\beta}$ and $n_\perp\approx1/\be$.

The Bohr-Sommerfeld quantization condition,
\begin{equation}
    4S(\bar{\theta}_0)=4k_0\s(\theta_0)=2\pi(n+1/2),\qquad n=\overline{0,\infty},
\end{equation}
gives rise to the spectrum
\begin{equation}\label{wkb_spectrum}
    k_{0n}=\frac{\pi q}{4}\frac{2n+1}{|\de\e|^{1/2}n_\perp\Phi(1/\vk)},\qquad\Phi(\tau):=E(\tau)+(\tau-1)K(\tau),
\end{equation}
where $E(\tau)$ and $K(\tau)$ are the complete elliptic integrals, and it has been used that
\begin{equation}
    \s(\bar{\theta}_0)=\frac{n_\perp}{q}|\de\e|^{1/2}\Phi(1/\vk).
\end{equation}
In making the estimates, one can put $\Phi(\tau)\approx \tau$ for $\tau\in[0,1]$ (see Fig. \ref{Phi_dP_plots}).

The semiclassical wave function \eqref{extraordinary_+} concentrated in the potential well near the point $\bar{\theta}=0$ is written as
\begin{equation}\label{psi_0}
    \psi_0(\bar{\theta}):= \Big(2k_3^{(2)}\bar{k}_0^2(\bar{k}_0^2-k_\perp^2\cos^2\bar{\theta})\Big)^{-1/2}
    \left[
      \begin{array}{c}
        \zeta^* \\
        \zeta \\
      \end{array}
    \right]e^{iS(\bar{\theta})},
\end{equation}
where $k_0=k_{0n}$, the principal branches of the multivalued functions are taken, and the replacement $\bar{\theta}\rightarrow\bar{\theta}-i0$, $\bar{\theta}\in \mathbb{R}$, is understood. Such a prescription for $\bar{\theta}$ ensures the exponential damping of the wave function under the potential barrier. We will refer to this branch of the multivalued function $\psi_0(\bar{\theta})$ as physical. The approximate Bloch wave function reads (see, e.g., \cite{LandLifshST2})
\begin{equation}\label{Bloch_sol}
    \psi(p_3;\bar{\theta})=\sum_{k=-\infty}^\infty \psi_{0\kappa}(\bar{\theta}-\pi k)e^{i\pi k p_3/q},
\end{equation}
where $p_3\in[-q,q)$ is the quasimomentum of the wave function and $\psi_{0\kappa}(\bar{\theta})$ has the form \eqref{psi_0} with the replacement $k_0\rightarrow k_0(p_3)$. The quasimomentum is an arbitrary parameter and it, of course, does not coincide with \eqref{quasimom_p3}.

Then, following \cite{LandLifshST2}, we derive the splitting of the spectrum \eqref{wkb_spectrum} into bands and the dispersion law $k_{0n}(p_3)$ of the photon in the CLC. Up to small terms in the shortwave approximation, we can write
\begin{equation}\label{Max_eqns_4}
    \big[q^2K\partial_{\bar{\theta}}^2+\bar{k}_0^2\bar{V}\big]\psi_0(\bar{\theta})=0,\qquad \big[q^2K\partial_{\bar{\theta}}^2+\bar{k}_0^2(p_3)\bar{V}\big]\psi(p_3;\bar{\theta})=0,
\end{equation}
where
\begin{equation}
    \bar{V}:=V/\bar{k}_0^2,\qquad \bar{k}_0^2(p_3):=k_0^2(p_3)\e_\perp.
\end{equation}
In Eqs. \eqref{Max_eqns_4}, we formally suppose that $k_0$ entering into the components of the permittivity tensor does not depend on  $k_{0n}$ or $k_0(p_3)$ and is the same for both the equations in \eqref{Max_eqns_4}. Then the short wavelength solutions of \eqref{Max_eqns_4} have the form \eqref{psi_0}, \eqref{Bloch_sol}, where one should regard $k_0$ in the components of the permittivity tensor as an independent quantity. Only in the final expression will we set $k_0$ to be equal to the physical value. It follows from Eqs. \eqref{Max_eqns_4} that
\begin{equation}\label{eqn_dispers}
    [k_0^2(p_3)-k_0^2]\e_\perp\int_{-\pi/2}^{\pi/2} d\bar{\theta} \psi^\dag(p_3;\bar{\theta})\bar{V} \psi_0(\bar{\theta})=q^2\Big[\psi^\dag(p_3;\bar{\theta})K\psi'_0(\bar{\theta}) -\psi'^\dag(p_3;\bar{\theta})K\psi_0(\bar{\theta})\Big]^{\bar{\theta}=\pi/2}_{\bar{\theta}=-\pi/2},
\end{equation}
where one should put $k_0=k_{0n}$ everywhere save the components $\e_{ij}(k_0)$, and the prime denotes the derivative with respect to $\bar{\theta}$. The expression \eqref{eqn_dispers} determines implicitly the dispersion relation $k_0(p_3)$.

Let us simplify Eq. \eqref{eqn_dispers} under the assumption that the band splitting of the levels $k_{0n}$ is small in comparison with the distance between the adjacent levels. Taking into account that
\begin{equation}
    S^*(\bar{\theta}-i0)=-S(-\bar{\theta}-i0),\qquad [k_3^{(2)}(\bar{\theta}-i0)]^*=k_3^{(2)}(-\bar{\theta}-i0),\qquad\zeta^*(\bar{\theta})=-\sgn(\de\e)\zeta(-\bar{\theta}),
\end{equation}
we have the relations
\begin{equation}\label{symm_prop}
    \psi_0^*(\bar{\theta})=-\sgn(\de\e) \psi_0(-\bar{\theta}),\qquad\psi_0'^*(\bar{\theta})=\sgn(\de\e) \psi'_0(-\bar{\theta}),
\end{equation}
on the physical branch of the wave functions for $\bar{\theta}\in \mathbb{R}$. Employing these relations, the right-hand side of  \eqref{eqn_dispers} can be rewritten as
\begin{equation}\label{RHS_eqn}
    -2q\im\Big[(k_{30}^{(2)}(\tfrac\pi2)+k_{3\kappa}^{(2)*}(\tfrac\pi2))\psi^\dag_{0\kappa}(\tfrac\pi2)K\psi_0(\tfrac\pi2) -\sgn(\de\e)(k_{30}^{(2)}(\tfrac\pi2)+k_{3\kappa}^{(2)}(\tfrac\pi2))\psi^T_{0\kappa}(\tfrac\pi2)K\psi_0(\tfrac\pi2)e^{-i\pi p_3/q} \Big],
\end{equation}
up to exponentially suppressed terms in the shortwave approximation. Here
\begin{equation}
    k_{30}^{(2)}(\tfrac\pi2)=ik_{0n}|\e_\parallel-n_\perp^2|^{1/2},\qquad k_{3\kappa}^{(2)}(\tfrac\pi2)=ik_0(p_3)|\e_\parallel-n_\perp^2|^{1/2},
\end{equation}
and
\begin{equation}
\begin{split}
    \psi_0(\tfrac\pi2)&=\bigg(\frac{i}{2k_{0n}|\e_\parallel-n_\perp^2|^{1/2}}\bigg)^{1/2}
    \left[
      \begin{array}{c}
        1 \\
        -1 \\
      \end{array}
    \right]
    e^{k_{0n}[i\s(\theta_0)-n_\perp|\de\e|^{1/2}\Phi(1-1/\vk)/q]},\\
    \psi_{0\kappa}(\tfrac\pi2)&=\bigg(\frac{i}{2k_{0}(p_3)|\e_\parallel-n_\perp^2|^{1/2}}\bigg)^{1/2}
    \left[
      \begin{array}{c}
        1 \\
        -1 \\
      \end{array}
    \right]
    e^{k_{0}(p_3)[i\s(\theta_0)-n_\perp|\de\e|^{1/2}\Phi(1-1/\vk)/q]}.
\end{split}
\end{equation}
Notice that
\begin{equation}
    1-1/\vk=\frac{\e_\parallel-n_\perp^2}{\de\e n_\perp^2}.
\end{equation}
As long as we suppose that $k_0(p_3)\approx k_{0n}$, the first term in \eqref{RHS_eqn} is much smaller than the second term in \eqref{RHS_eqn}. Therefore, in the leading order in
\begin{equation}
    \de k_0:=k_0(p_3)-k_{0n},
\end{equation}
one can neglect the first contribution in \eqref{RHS_eqn} as compared with the second one.

Now we perform the integral on the left-hand side of \eqref{eqn_dispers}. It is readily found that the integrand
\begin{equation}
    \psi^\dag_{0\kappa}\bar{V}\psi_0=\frac{1}{\e_\perp\sqrt{k_{0n}k_{0}(p_3)}} \frac{\big[n_3^{(2)}(\bar{\theta})\big]^2}{|n_3^{(2)}(\bar{\theta})|}e^{ik_{0n}\s(\bar{\theta})-ik_{0}(p_3)\s^*(\bar{\theta})},
\end{equation}
where
\begin{equation}
    n_3^{(2)}(\bar{\theta}):=\frac{k_{30}^{(2)}(\bar{\theta})}{k_{0n}}=q\s'(\bar{\theta}).
\end{equation}
Then we split the integral on the left-hand side of \eqref{eqn_dispers} in the following way
\begin{equation}
    \bigg(\int_{-\pi/2}^{-\bar{\theta_0}}d\bar{\theta} +\int_{-\bar{\theta_0}}^{\bar{\theta_0}}d\bar{\theta} +\int_{\bar{\theta_0}}^{\pi/2}d\bar{\theta}\bigg)\psi^\dag_{0\kappa}\bar{V}\psi_0=:I_1+I_2+I_3.
\end{equation}
The symmetry properties imply
\begin{equation}
    I_1=I_3^*.
\end{equation}
The integrals are evaluated by parts
\begin{equation}
    I_2=\frac{2q}{\e_\perp\sqrt{k_{0n}k_{0}(p_3)}}\frac{\sin[(k_{0n}-k_{0}(p_3))\s(\bar{\theta}_0)]}{k_{0n}-k_{0}(p_3)},\qquad I_3\approx -\frac{q}{\e_\perp\sqrt{k_{0n}k_{0}(p_3)}}\frac{e^{i(k_{0n}-k_{0}(p_3))\s(\bar{\theta}_0)}}{k_{0n}+k_{0}(p_3)}.
\end{equation}
The upper limit in the last integral was put to $+\infty$. This is justified up to negligible contributions in the shortwave approximation. As a result,
\begin{equation}
    I_1+I_2+I_3=\frac{2q}{\e_\perp\sqrt{k_{0n}k_{0}(p_3)}}\frac{\sin[(k_{0n}-k_{0}(p_3))\s(\bar{\theta}_0)]}{k_{0n}-k_{0}(p_3)} -\frac{2q}{\e_\perp\sqrt{k_{0n}k_{0}(p_3)}}\frac{\cos[(k_{0n}-k_{0}(p_3))\s(\bar{\theta}_0)]}{k_{0n}+k_{0}(p_3)}.
\end{equation}
For $|\de k_0\s(\bar{\theta}_0)|\ll1$ and $k_{0n}\s(\bar{\theta}_0)\gg1$, the second term in this expression can be neglected. The condition $|\de k_0\s(\bar{\theta}_0)|\ll1$ means that the width of the band is much smaller than the distance between the bands.

Setting $k_0=k_{0n}$ in the components of the permittivity tensor, Eq. \eqref{eqn_dispers} turns into
\begin{equation}
    \de k_0\s(\bar{\theta}_0)=\sin[2k_{0n}\s(\bar{\theta}_0)-\pi p_3/q]e^{-2k_\perp|\de\e|^{1/2}\Phi(1-1/\vk)/q},
\end{equation}
in the leading order in $\de k_0$. Substituting the spectrum \eqref{wkb_spectrum}, we arrive at the dispersion law
\begin{equation}\label{disper_law}
    k_0(p_3)=k_{0n}\Big(1+\frac{4}{\pi}\frac{(-1)^n}{2n+1}\cos\frac{\pi p_3}{q}e^{-2k_\perp|\de\e|^{1/2}\Phi(1-1/\vk)/q}\Big).
\end{equation}
As is seen, the spectrum \eqref{wkb_spectrum} is split into the bands
\begin{equation}\label{bands}
    k_0(p_3)\in k_{0n}[1-\frac{4}{\pi}\frac{e^{-2k_\perp|\de\e|^{1/2}\Phi(1-1/\vk)/q}}{2n+1}, 1+\frac{4}{\pi}\frac{e^{-2k_\perp|\de\e|^{1/2}\Phi(1-1/\vk)/q}}{2n+1}],\qquad n=\overline{0,\infty}.
\end{equation}
The velocity of the wave packet along the $z$ axis is evidently
\begin{equation}
    \ups_3:=\frac{\partial k_0(p_3)}{\partial p_3}= \frac{4 k_{0n}}{q}\frac{(-1)^{n+1}}{2n+1}\sin\frac{\pi p_3}{q}e^{-2k_\perp|\de\e|^{1/2}\Phi(1-1/\vk)/q}.
\end{equation}
In the case when the tunneling is small and so the exponential factor is tiny, the velocity of the wave packet of a photon along the $z$ axis is also very small. For $|p_3|\ll q$, the dispersion relation with respect to $p_3$ for fixed $n_\perp$ takes the form of the dispersion law of a nonrelativistic particle with the effective mass
\begin{equation}\label{mass_eff}
    m_{eff}=(-1)^{n+1}\frac{q^2(n+1/2)}{2\pi k_{0n}} e^{2k_\perp|\de\e|^{1/2}\Phi(1-1/\vk)/q}.
\end{equation}
The effective mass is negative when $n$ is even.

Consider the case $\de\e>0$. For the minimum of the potential well to be located at $\bar{\theta}=0$, we make the replacement $\bar{\theta}\rightarrow \bar{\theta}-\pi/2$. Then the Hamilton-Jacobi action becomes
\begin{equation}
    S(\bar{\theta})=\frac{1}{q}\int_0^{\bar{\theta}}d\tau k_3^{(2)}(\tau)=\frac{k_0}{q}\int_0^{\bar{\theta}} d\tau\sqrt{(1+\de\e)\e_\perp-n_\perp^2-\de\e n_\perp^2\sin^2\tau}=:k_0\s(\bar{\theta}).
\end{equation}
The turning point has the form \eqref{turn_point} with
\begin{equation}
    \vk=\frac{\de\e n_\perp^2}{\e_\parallel-n_\perp^2}.
\end{equation}
This turning point appears only when
\begin{equation}\label{np_dep}
    \e_\perp\leqslant n_\perp^2\leqslant \e_\parallel.
\end{equation}
As in the case $\de\e<0$, the modes under study tunnel through the CLC plate when $n_\perp<1$. This may happen only when $\e_\perp<1$. For $n_\perp>1$, these modes are left in the CLC plate. As for the ordinary wave, it does not propagate through the CLC plate when the conditions \eqref{np_dep} are met. The Bohr-Sommerfeld quantization condition results in the spectrum \eqref{wkb_spectrum}.

The semiclassical wave function concentrated in the potential well near the point $\bar{\theta}=0$ is given by
\begin{equation}
    \psi_0(\bar{\theta}):= \Big(2k_3^{(2)}\bar{k}_0^2(\bar{k}_0^2-k_\perp^2\sin^2\bar{\theta})\Big)^{-1/2}
    \left[
      \begin{array}{c}
        \zeta^* \\
        \zeta \\
      \end{array}
    \right]e^{iS(\bar{\theta})},
\end{equation}
where
\begin{equation}
    \zeta=\bar{k}_3^2\sin\bar{\theta}+i\bar{k}_0^2\cos\bar{\theta}.
\end{equation}
The physical branch of the wave function is specified as in the case $\de\e<0$ considered above. The Bloch wave function is constructed in accordance with the formula \eqref{Bloch_sol}. The splitting of the spectrum into bands is found analogously to what was done in the case $\de\e<0$, but the variable $\bar{\theta}$ in the potential $\bar{V}$ should be shifted as $\bar{\theta}\rightarrow\bar{\theta}-\pi/2$. Then the formulas \eqref{eqn_dispers}-\eqref{RHS_eqn} are valid with
\begin{equation}
    k_{30}^{(2)}(\tfrac\pi2)=ik_{0n}\sqrt{1+\de\e}|\e_\perp-n_\perp^2|^{1/2},\qquad k_{3\kappa}^{(2)}(\tfrac\pi2)=ik_0(p_3)\sqrt{1+\de\e}|\e_\perp-n_\perp^2|^{1/2},
\end{equation}
and
\begin{equation}
\begin{split}
    \psi_0(\tfrac\pi2)&=-\bigg(\frac{i|\e_\perp-n_\perp^2|^{1/2}}{2k_{0n}\e_\perp\sqrt{1+\de\e}}\bigg)^{1/2}
    \left[
      \begin{array}{c}
        1 \\
        1 \\
      \end{array}
    \right]
    e^{k_{0n}[i\s(\theta_0)-n_\perp|\de\e|^{1/2}\Phi(1-1/\vk)/q]},\\
    \psi_{0\kappa}(\tfrac\pi2)&=-\bigg(\frac{i|\e_\perp-n_\perp^2|^{1/2}}{2k_0(p_3)\e_\perp\sqrt{1+\de\e}}\bigg)^{1/2}
    \left[
      \begin{array}{c}
        1 \\
        1 \\
      \end{array}
    \right]
    e^{k_0(p_3)[i\s(\theta_0)-n_\perp|\de\e|^{1/2}\Phi(1-1/\vk)/q]}.
\end{split}
\end{equation}
Besides,
\begin{equation}
    1-1/\vk=(1+\de\e)\frac{n_\perp^2-\e_\perp}{\de\e n_\perp^2}.
\end{equation}
The scalar product entering into the integrand of the integral in \eqref{eqn_dispers} becomes
\begin{equation}
    \psi^\dag_{0\kappa}\bar{V}\psi_0=\frac{\sgn(\e_\perp-n_\perp^2\sin^2\bar{\theta})}{\e_\perp\sqrt{k_{0n}k_{0}(p_3)}} \frac{\big[n_3^{(2)}(\bar{\theta})\big]^2}{|n_3^{(2)}(\bar{\theta})|}e^{ik_{0n}\s(\bar{\theta})-ik_{0}(p_3)\s^*(\bar{\theta})}.
\end{equation}
The integration of this expression over $\bar{\theta}$ can be performed by parts. Then, carrying out the calculations as in the case $\de\e<0$, it is not difficult to deduce that, in the leading order in $\de k_0$, the dispersion relation has the form \eqref{disper_law} up to exponentially suppressed terms in the shortwave approximation.

Thus we see that, for $n_\perp<1$ belonging to the interval \eqref{np_dem} or \eqref{np_dep}, the medium with permittivity  \eqref{permit_holec} transmits extraordinary electromagnetic waves with energies \eqref{bands} and can be used as a filter in this energy range provided $\ups_3$ is not very small. In the case \eqref{np_dep}, the ordinary waves do not penetrate through the CLC layer. Electromagnetic waves with energies out of the range \eqref{bands} are completely reflected from the plate made of such a medium. The same effect takes place for a plane-periodic isotropic medium with permittivity $\e(k_0;z)$ less than unity for some $z$. If
\begin{equation}
    \min\e\leqslant n_\perp^2\leqslant \max\e,
\end{equation}
there exists a band spectrum and so does a selective transmissivity of the electromagnetic waves by such a medium \cite{Sakoda,JJWMbook,IstSar06}.

Let us find the approximate dispersion relation with respect to $k_\perp$ for the bound modes in the case  when the dependence of $\e_{\perp,\parallel}$ on $k_0$ is marginal. Taking $\Phi(\tau)\approx\tau$ in \eqref{wkb_spectrum}, we come to
\begin{equation}
    k_0^2=\frac{k_\perp^2}{\max(\e_\perp,\e_\parallel)}+\frac{\pi q}{2}(n+1/2)|\de\e|^{1/2}\frac{k_\perp}{\e_\parallel},
\end{equation}
where
\begin{equation}
    k_\perp\in k_0\big[\min(\e_\perp^{1/2},\e_\parallel^{1/2}),\max(\e_\perp^{1/2},\e_\parallel^{1/2})\big],
\end{equation}
and, of course, the condition of applicability of the shortwave approximation \eqref{short_wave2} should be satisfied. The dependence of $\e_{\perp,\parallel}$ on $k_0$ can easily be taken into account regarding the varying part of $\e_{\perp,\parallel}(k_0)$ as a small perturbation.


\subsection{Plasma permittivity}\label{Plasm_Perm}

The case when the permittivity tensor has the form \eqref{permit_UV} is of particular importance. The components of the permittivity tensor have such a form for plasmonic metamaterials \cite{Sakoda,JJWMbook,PiSiChuEch} and for any medium at sufficiently high energies of photons. As we have discussed in Sec. \ref{Max_Eqs}, at high photon energies the CLCs consisting of rod-shaped molecules possess the permittivity tensor of the form \eqref{permit_UV} with $\de\e<0$, whereas $\de\e>0$ for the CLCs made of disc-shaped molecules. The both signs of $\de\e$ are realized in plasmonic metamaterials.

\begin{figure}[t]
   \centering
   \includegraphics*[width=0.49\linewidth]{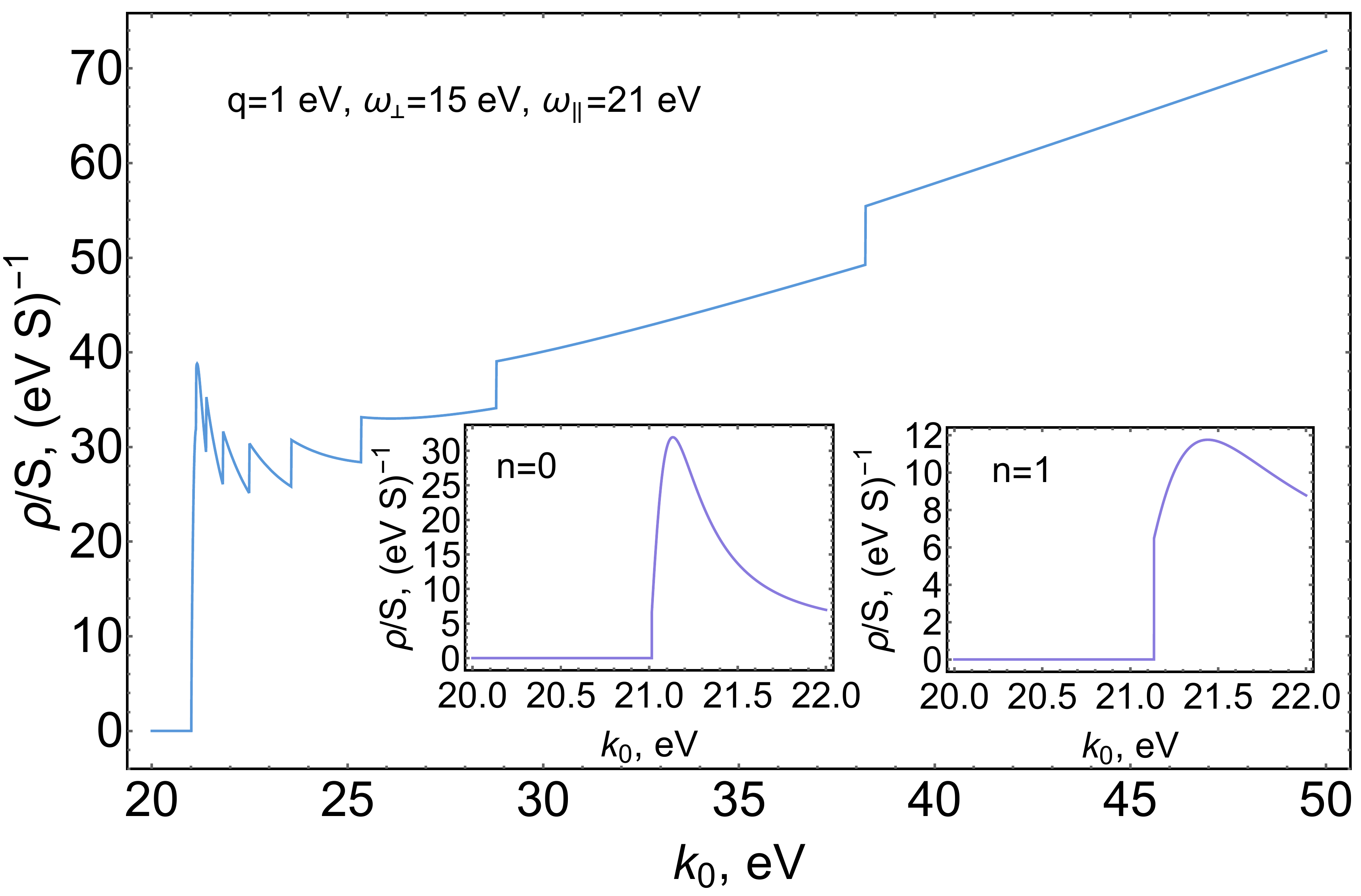}\;
   \includegraphics*[width=0.49\linewidth]{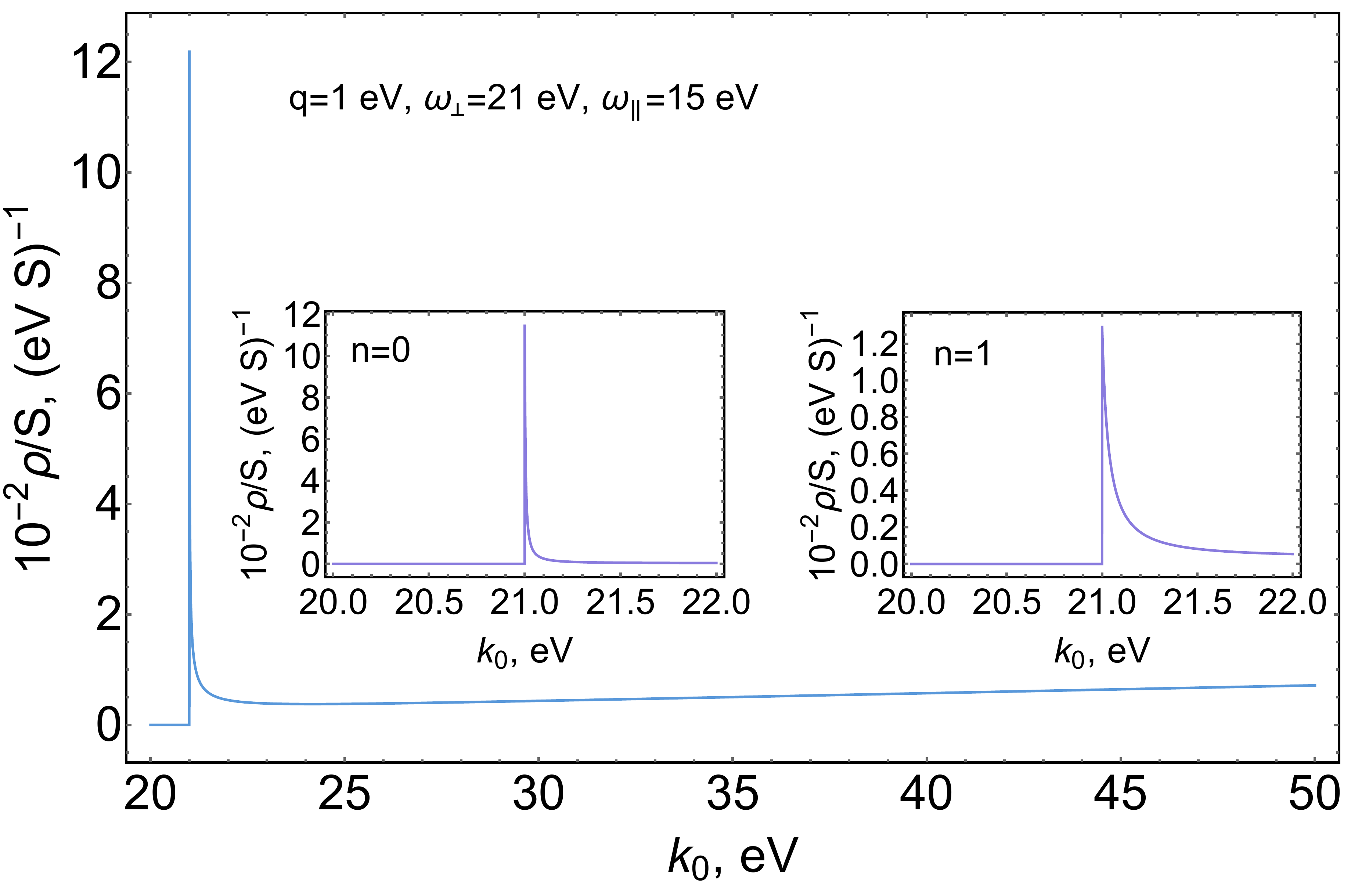}\;
   \caption{{\footnotesize The contribution to the density of states from the bound states. The insets depict the contributions to the density of states from the separate bound states. On the left panel: The case $\de\e<0$ is plotted. On the right panel: The case $\de\e>0$ is presented.}}
\label{DOS_plots}
\end{figure}

Replacing $\Phi(\tau)\approx\tau$ in the spectrum \eqref{wkb_spectrum}, we deduce that in the case $\de\e<0$ the spectrum of bound states is implicitly determined by the equation
\begin{equation}\label{spectrum_plm_dem}
    \frac{2}{\pi q}\frac{(k_0^2-\omega_{\parallel}^2)(k_3^2-\omega_\perp^2)}{k_\perp\sqrt{(\omega_\parallel^2-\omega_\perp^2) (k_0^2-\omega_\perp^2)}} =n+1/2,
\end{equation}
where
\begin{equation}\label{k0_int_dem}
    \max(\omega_\parallel,\omega_\perp/n_3)\leqslant k_0 \leqslant \omega_\parallel/n_3,\qquad n_3=\sqrt{1-n_\perp^2}.
\end{equation}
For the given interval of $k_0$, the function on the left-hand side of \eqref{spectrum_plm_dem} is monotonically increasing and reaches the maximum at $k_0=\omega_\parallel/n_3$. Consequently, the bound states are realized in such a medium only when
\begin{equation}
    \frac{4\omega_\parallel n_\perp}{\pi q}\sqrt{\frac{\omega_\parallel^2-\omega_\perp^2}{\omega_\parallel^2-\omega_\perp^2 n_3^2}}>1.
\end{equation}
If $\de\e>0$ for the components of the permittivity tensor \eqref{permit_UV}, then the spectrum is found from
\begin{equation}\label{spectrum_plm_dep}
    \frac{2}{\pi qk_\perp}\sqrt{\frac{k_0^2-\omega_{\perp}^2}{\omega_\perp^2-\omega_\parallel^2}} (k_3^2-\omega_\parallel^2) =n+1/2,
\end{equation}
where
\begin{equation}\label{k0_int_dep}
    \max(\omega_\perp,\omega_\parallel/n_3)\leqslant k_0 \leqslant \omega_\perp/n_3.
\end{equation}
As in the case $\de\e<0$, this function is monotonically increasing on the interval \eqref{k0_int_dep} and reaches the maximum at $k_0=\omega_\perp/n_3$. Then the necessary condition for the existence of bound states reads
\begin{equation}
    \frac{4}{\pi q}\sqrt{\omega_\perp^2-\omega_\parallel^2}>1.
\end{equation}
In the case when $\omega_{\perp,\parallel}^2/k_0^2\ll1$, the spectrum can explicitly be found. For the both signs of $\de\e$, we obtain
\begin{equation}\label{wkb-spectrum_plam}
    k^2_0-k_\perp^2\approx\min(\omega_\perp^2,\omega_\parallel^2) +\frac{\pi q}{2} (n+1/2) \big|\omega_{\parallel}^2-\omega_{\perp}^2\big|^{1/2},
\end{equation}
where the photon energy is bounded by inequalities \eqref{k0_int_dem} or \eqref{k0_int_dep}. The dispersion law \eqref{wkb-spectrum_plam} has the form of the dispersion relation of a massive relativistic particle in a three-dimensional spacetime. The right-hand side of \eqref{wkb-spectrum_plam} is the corresponding mass squared.

Since $\e_{\parallel,\perp}<1$ for the plasma permittivity, inequalities \eqref{np_dem}, \eqref{np_dep} imply that $n_\perp<1$. Therefore, the states considered are not bound in the CLC plate of a finite width as they tunnel through the potential barrier to the vacuum. If one takes into account this process, then these states become resonances with the lifetime inversely proportional to the tunneling probability. The presence of such resonances results in a rapid increase of the transmittance coefficient and the density of states in their neighborhood (see \eqref{dens_of_states}). The increase of the density of states, in turn, leads to an amplification of quantum optical effects in the CLC plate \cite{Sakoda}.

Let us find the corresponding contribution to the density of states from a single state belonging to the spectrum \eqref{spectrum_plm_dem}, \eqref{spectrum_plm_dep}. Introduce for brevity
\begin{equation}
    g^-_n:=\frac{\pi q}{4}(n+1/2)\frac{\sqrt{(\omega_\parallel^2-\omega_\perp^2)(k_0^2-\omega_\perp^2)}}{k_0^2-\omega_\parallel^2},\qquad g^+_n:=\frac{\pi q}{4}(n+1/2)\sqrt{\frac{\omega_\perp^2-\omega_\parallel^2}{k_0^2-\omega_\perp^2}}.
\end{equation}
Then we infer from \eqref{spectrum_plm_dem}, \eqref{spectrum_plm_dep} that
\begin{equation}\label{disper_law_bound}
    k_\perp=\sqrt{(g_n^\pm)^2+k_0^2-\omega_{\perp,\parallel}^2}-g^\pm_n,
\end{equation}
where $g^+_n$ and $\omega_\parallel$ are taken in the case $\de\e>0$, whereas $g^-_n$ and $\omega_\perp$ are taken in the case $\de\e<0$. The contribution to the density of states becomes
\begin{equation}
    \rho(k_0)=\int\frac{Sd\spk_\perp}{(2\pi)^2}\de(k_0-k_0(k_\perp))=\frac{S}{2\pi}\frac{dk_\perp}{dk_0}k_\perp(k_0),
\end{equation}
where $S$ is the area of the CLC plate and the inequality \eqref{k0_int_dem} or $\eqref{k0_int_dep}$ should be fulfilled for the respective sign of $\de\e$.

\begin{figure}[t]
   \centering
   \includegraphics*[width=0.49\linewidth]{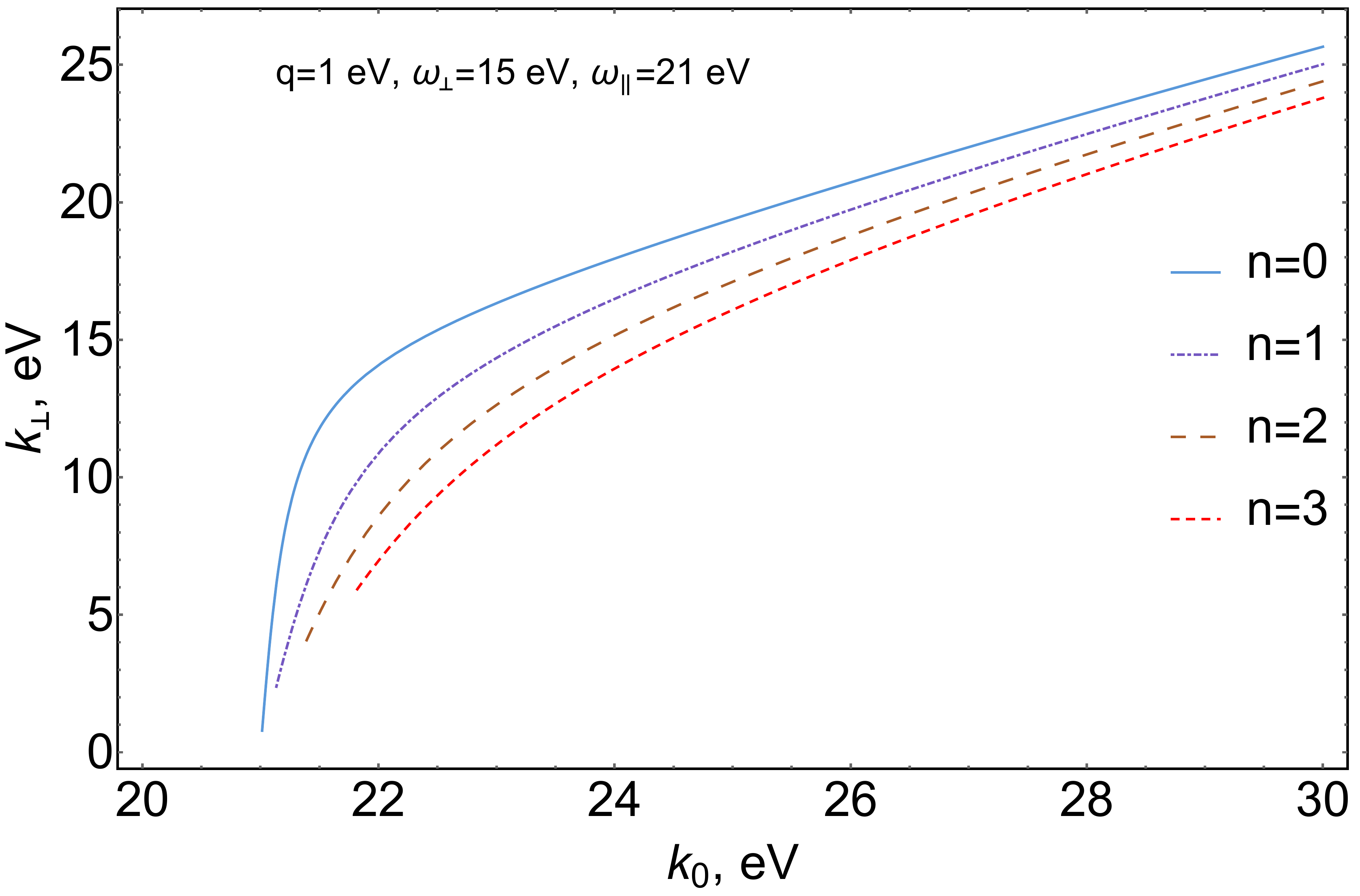}\;
   \includegraphics*[width=0.49\linewidth]{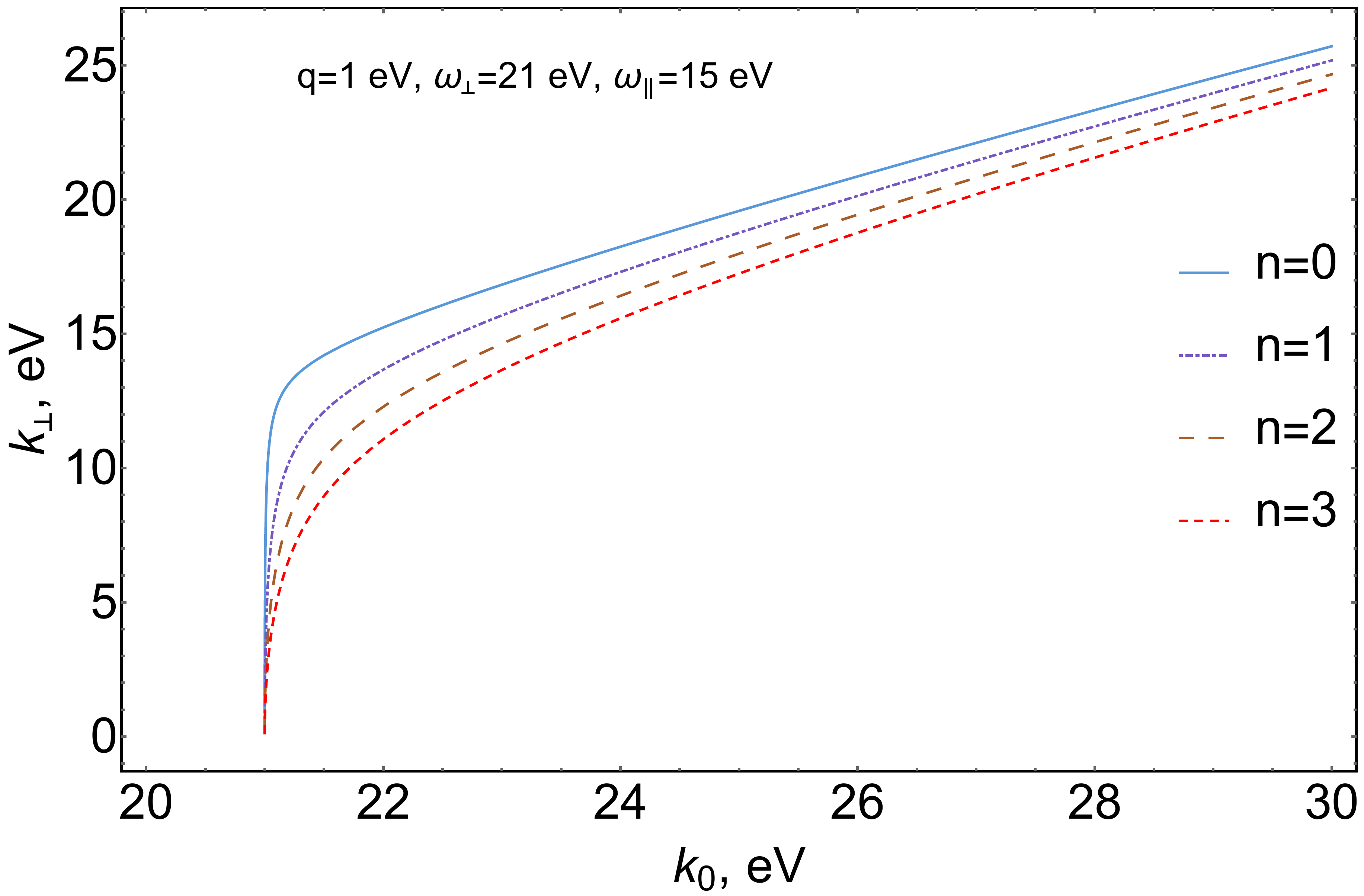}\;
   \caption{{\footnotesize The dispersion laws of photons in the bound states. The case $\de\e<0$ is on the left panel. The case $\de\e>0$ is on the right panel.}}
\label{Disp_Rels_plots}
\end{figure}

As expected, for large energies $\rho(k_0)\approx Sk_0/(2\pi)$ since the condition \eqref{wkb-spectrum_plam} fixes the third momentum component. If $q/\omega_{\perp,\parallel}\ll1$, then in the case $\de\e<0$ the spectral density possesses a sharp peak in a small vicinity of the point $k_0=\omega_\parallel$,
\begin{equation}
    \frac{\rho}{S}\approx \frac{8\omega_\parallel^2 (k_0-\omega_\parallel)}{\pi^3q^2(n+1/2)^2} +\frac{12\omega_\parallel (3\omega_\parallel^2-\omega_\perp^2)(k_0-\omega_\parallel)^2}{\pi^3q^2(n+1/2)^2(\omega_\parallel^2-\omega_\perp^2)} -\frac{512\omega_\parallel^4-4\pi^2q^2(n+1/2)^2(13\omega_\parallel^2-\omega_\perp^2)}{\pi^5q^4(n+1/2)^4 (\omega_\parallel^2-\omega_\perp^2)} (k_0-\omega_\parallel)^3,
\end{equation}
where inequality \eqref{k0_int_dem} must also be taken into account. For $\de\e>0$ such a peak is located in a small vicinity of the point $k_0=\omega_\perp$ where
\begin{equation}
    \frac{\rho}{S}\approx\frac{2\omega_\perp(\omega_\perp^2-\omega_\parallel^2)}{\pi^3q^2 (n+1/2)^2} -\frac{128\omega_\perp^2(\omega_\perp^2-\omega_\parallel^2) -4\pi^2q^2(n+1/2)^2(9\omega_\perp^2-\omega_\parallel^2)}{2\pi^5q^4(n+1/2)^4}(k_0-\omega_\perp).
\end{equation}
The plots of the density of states are presented in Fig. \ref{DOS_plots}. In the case $\de\e<0$, we have from \eqref{spectrum_plm_dem}, \eqref{k0_int_dem} in a small neighborhood of the point $k_0=\omega_\parallel$ that
\begin{equation}\label{lin_disp_law}
    k_0\approx\omega_\parallel+\frac{\pi q}{4\omega_\parallel}(n+1/2)k_\perp,\qquad k_\perp\geqslant\frac{\pi q}{2}(n+1/2).
\end{equation}
The photons obey a linear dispersion law along the plane of the CLC plate. When $\de\e>0$, Eq. \eqref{spectrum_plm_dep} and conditions \eqref{k0_int_dep} give rise to a quadratic dispersion relation,
\begin{equation}\label{quadr_disp_law}
    k_0\approx\omega_\perp +\frac{\pi^2q^2(n+1/2)^2}{8\omega_\perp(\omega_\perp^2-\omega_\parallel^2)}k_\perp^2,
\end{equation}
in the vicinity of the point $k_0=\omega_\perp$. As a rule, the effective mass is large for small $n$. The plots of the dispersion laws are given in Fig. \ref{Disp_Rels_plots}. As we saw, in the both cases $\de\e>0$ and $\de\e<0$, the component $\ups_3$ for the states at issue is exponentially suppressed for small $n$. Therefore, these modes propagate almost perpendicular to the CLC axis, i.e., almost parallel to the surface of the CLC plate.

\section{Resonances}\label{Reson_Sec}

The bound states and resonances in the CLC plate of a finite width can be found from the general condition \eqref{resonanses} or \eqref{resonanses1}. Comparing expressions \eqref{mode_funcz>0}-\eqref{mode_funcz<mL} with formula \eqref{2bases}, wee see that
\begin{equation}\label{alpha_k}
    \al^T(k_3)=
    \left[
      \begin{array}{cccc}
        1 & 0 & 0 & 0 \\
        0 & 1 & 0 & 0 \\
      \end{array}
    \right]
    (T')^{-1}H^{-1}UTU^{-1}
    \left[
      \begin{array}{cc}
        g(1) & g(-1) \\
      \end{array}
    \right],
\end{equation}
where $g(s)$ is defined in \eqref{joining_eqn_3}. Then
\begin{equation}\label{det_al}
\begin{split}
    \det\al(k_3)=&\,-\frac{e^{i(p_3+\bar{k}_3)L}}{4\e_\perp k_3^{(2)} k_3^2\bar{k}_3^2}\big[(k_3^2+\bar{k}_3^2)\sin(\bar{k}_3L)+2ik_3\bar{k}_3\cos(\bar{k}_3L)\big]\times\\
    &\times\big[\big(\bar{k}_3^4+\e_\perp^2 (k_3^{(2)})^2 k_3^2 \big)\sin(p_3L)+2i\e_\perp k_3^{(2)} k_3 \bar{k}_3^2\cos(p_3L)\big],
\end{split}
\end{equation}
where
\begin{equation}
    k_3^{(2)}=\sqrt{\e_\parallel k_0^2-(1+\de\e\cos^2\vf)k_\perp^2}.
\end{equation}
All the square roots in \eqref{det_al} in the definition of momenta are taken on the physical sheet with the branch cut along the positive real semiaxis, i.e., the branch with $\arg z\in[0,2\pi)$ is chosen in their definition. On the unphysical sheet,
\begin{equation}
    k_3=-\sqrt{k_0^2-k_\perp^2},
\end{equation}
and the branches of the roots entering into the definition of other momenta do not change.

\begin{figure}[t]
   \centering
\begin{tabular}{cc}
\includegraphics*[width=0.47\linewidth]{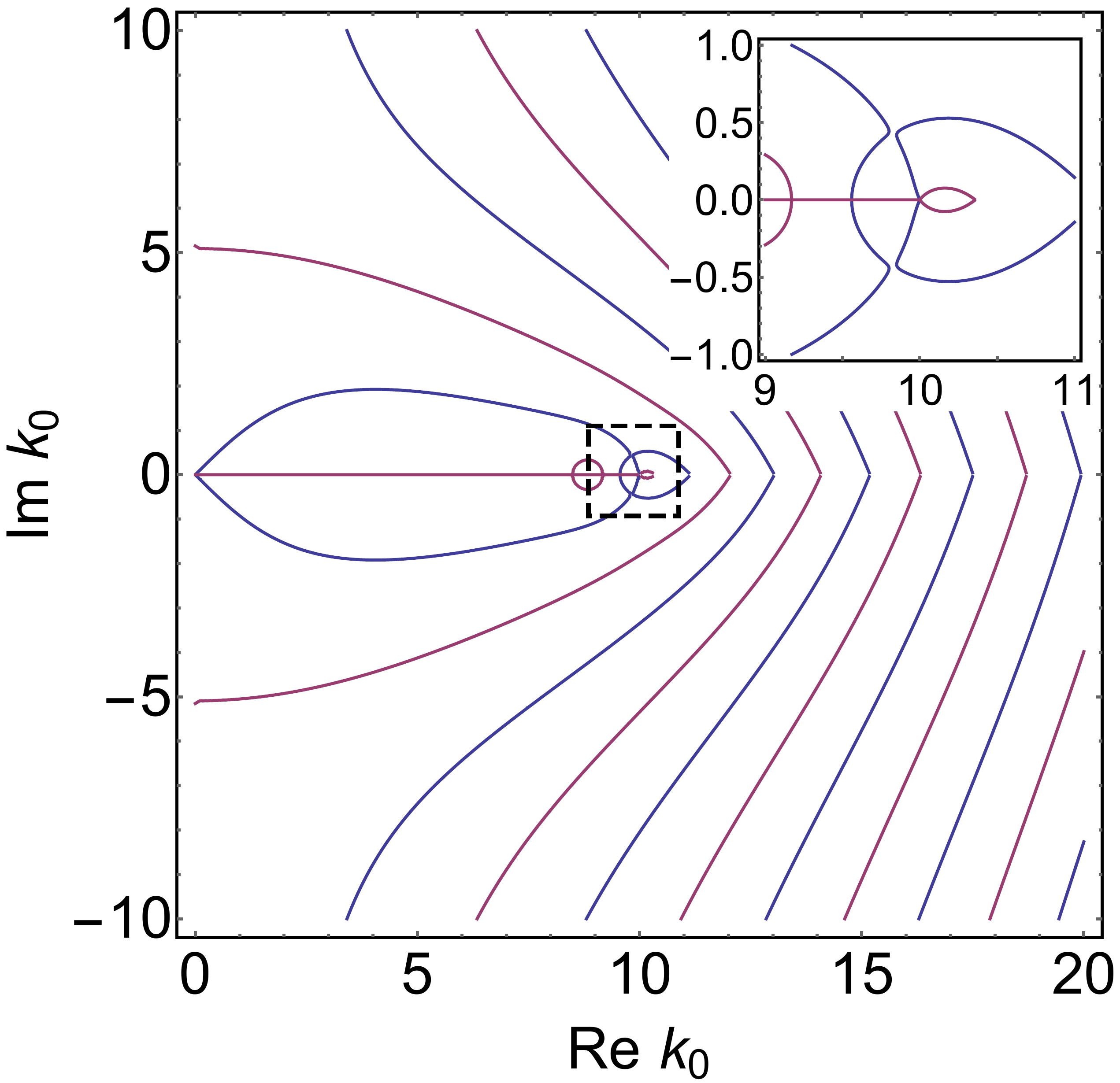}&
\includegraphics*[width=0.47\linewidth]{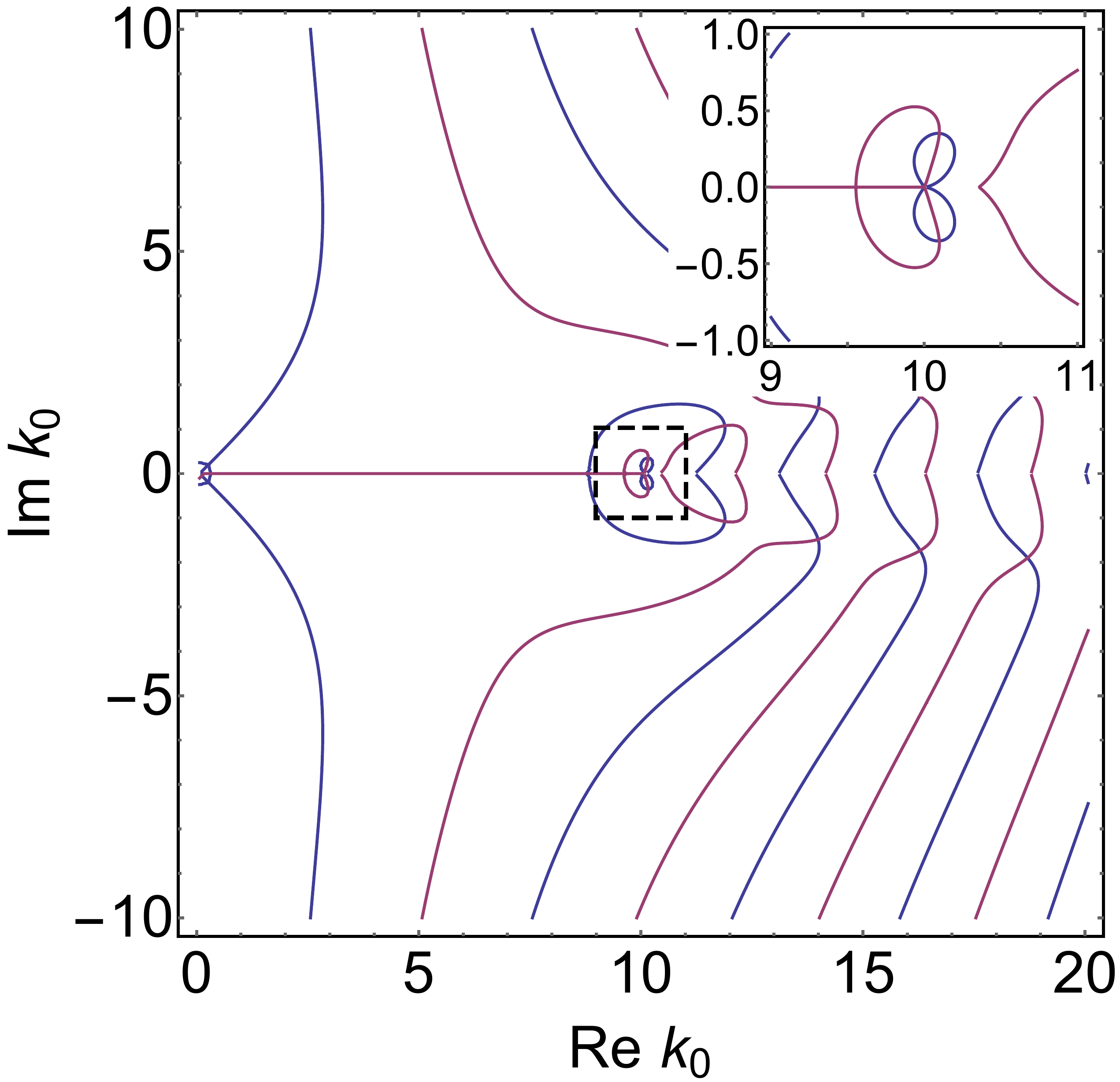}\\
\includegraphics*[width=0.47\linewidth]{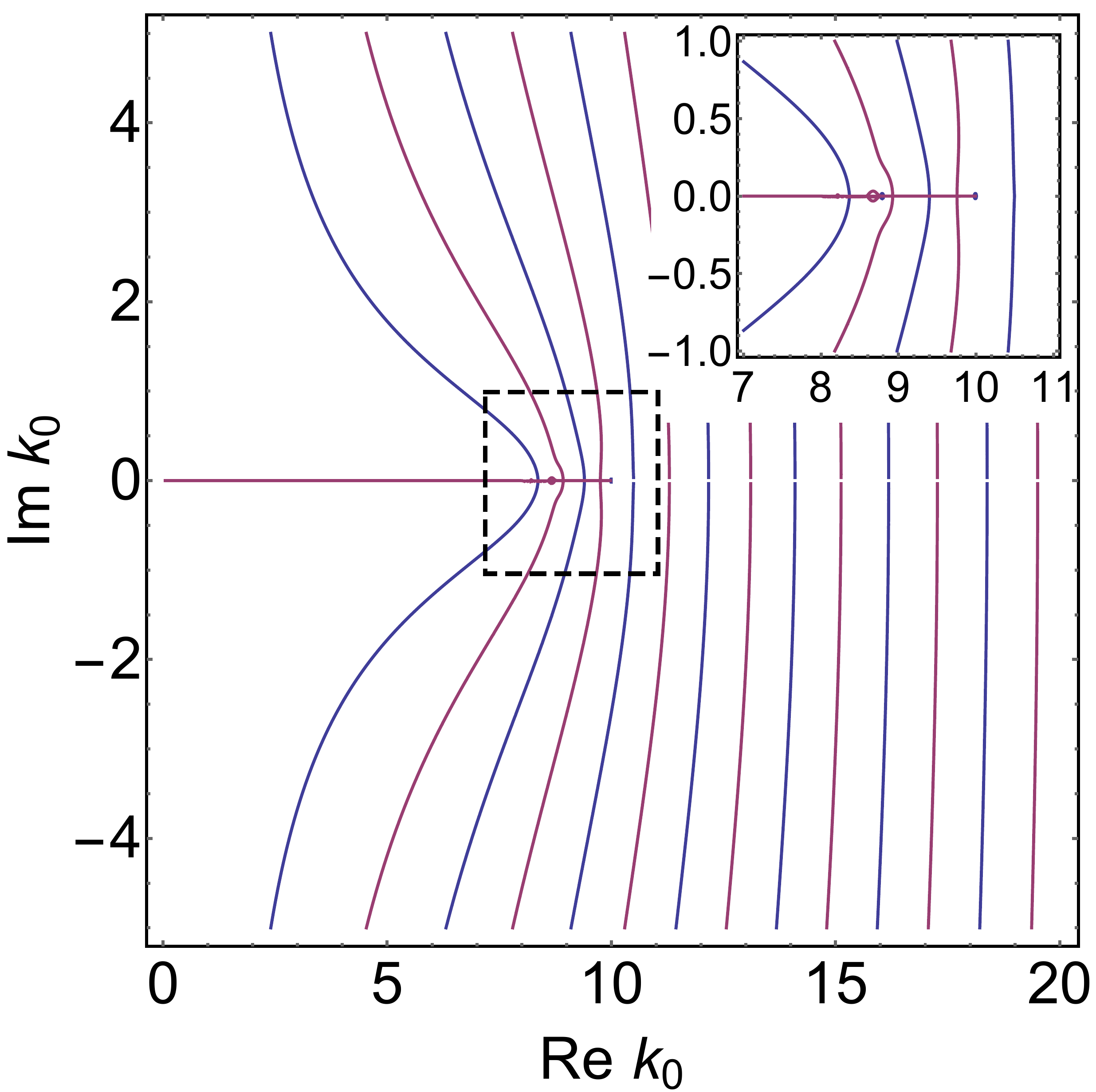}&
\includegraphics*[width=0.47 \linewidth]{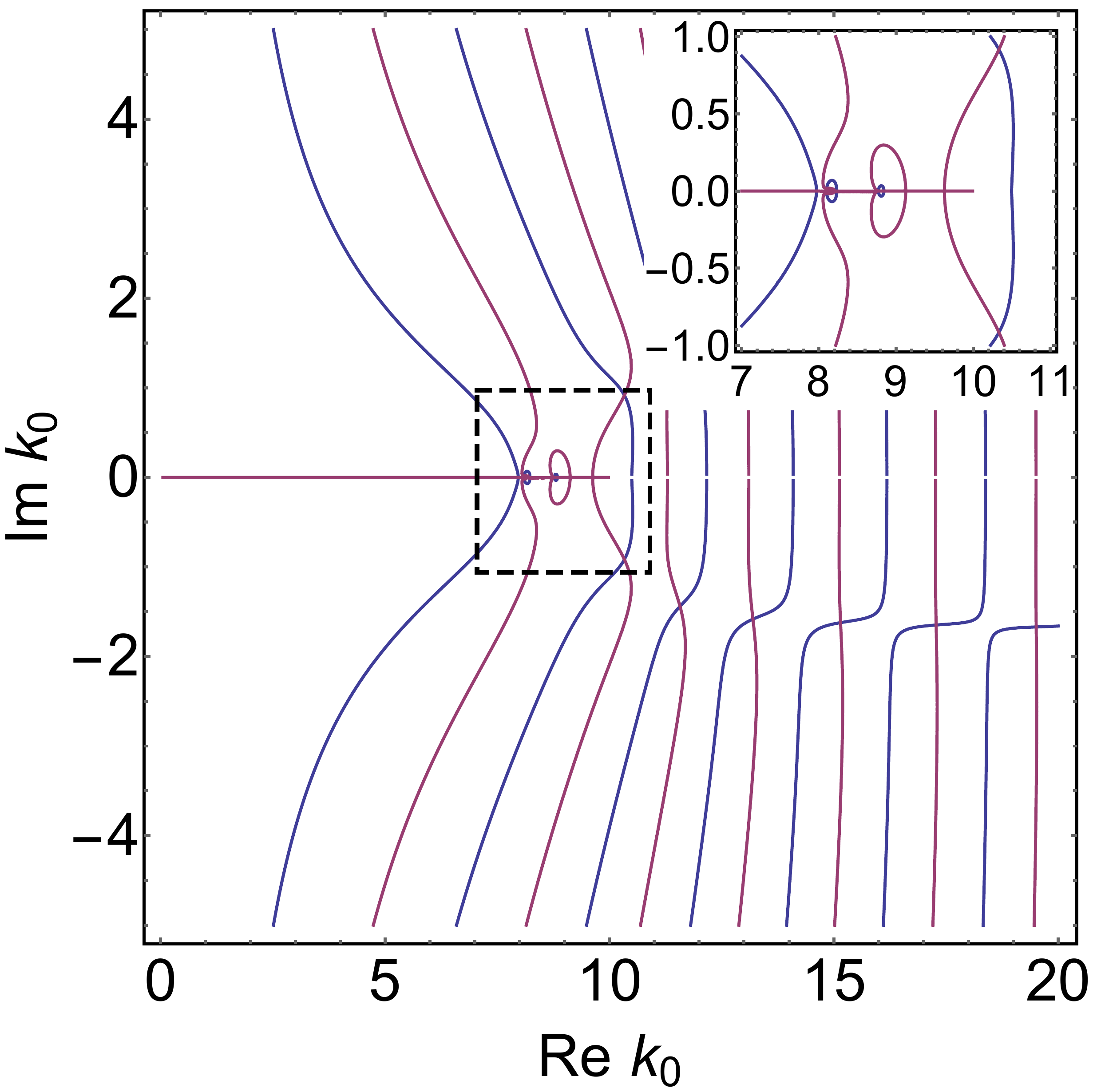}\\
\end{tabular}
    \caption{{\footnotesize The graphical solution of Eqs. \eqref{res_chol} for the bound states (on the left) and the resonances (on the right). The left plots corresponds to the physical sheet, $\im k_3\geqslant0$, whereas the right plots are for the unphysical one, $\im k_3<0$. The upper plots: The bound states and resonances of the ordinary wave in the CLC plate. The parameters are as follows $\e_\perp=1.3$, $L=1$, and $k_\perp=10$. The lines of different colors on the plots represent the points where the real or imaginary parts of the left-hand side of the first equation in \eqref{res_chol} vanish. The intersections of the lines of different colors correspond to the solutions of the first equation in \eqref{res_chol}. In particular, the bound states are the intersection points of lines of different colors lying on the segment [0,10] of the $\re k_0$ axis on the left plot. The lower plots: The bound states and resonances of the extraordinary wave in the CLC plate. The parameters are as follows $\e_\perp=1.3$, $\de\e=0.2$, $L=1$, $\vf=\pi/3$, and $k_\perp=10$. The lines of different colors on the plots represent the points where the real or imaginary parts of the left-hand side of the second equation in \eqref{res_chol} vanish. The intersections of the lines of different colors correspond to the solutions of the second equation in \eqref{res_chol}.}}
\label{Resons_plots}
\end{figure}

The condition \eqref{resonanses} is equivalent to
\begin{equation}\label{res_chol}
    \ctg (\bar{k}_3L)=\frac{i}{2}\Big(\frac{k_3}{\bar{k}_3} +\frac{\bar{k}_3}{k_3} \Big),\qquad \ctg (p_3L)=\frac{i}{2}\Big(\frac{\e_\perp k_3 k_3^{(2)}}{\bar{k}_3^2} +\frac{\bar{k}_3^2}{\e_\perp k_3 k_3^{(2)}} \Big).
\end{equation}
These equations are analogous to the equation for the bound states in the plate made of an isotropic homogeneous dielectric (see, e.g., (87) of \cite{BKL5}). The first equation describing the bound states of the ordinary wave coincides with the first equation in (87) of \cite{BKL5}. Similar equations arise in the analysis of the one-dimensional scattering problem of a non-relativistic particle on the potential well and in the analysis of the spectrum of surface polaritons in a thin plate \cite{PiSiChuEch}. A thorough description of solutions of the equation for  resonances in this case can be found in \cite{Nussen59,ZavMois04}. The equations \eqref{res_chol} cannot be solved exactly with respect to $k_0$. Their graphical solution is given in Fig. \ref{Resons_plots}.

Let us find the approximate solutions of Eqs. \eqref{res_chol}. We begin with the first equation. Suppose that $|k_\perp/k_0|\ll1$. Then the first equation in \eqref{res_chol} is approximately satisfied when (cf. \cite{BelyakovBook})
\begin{equation}\label{res1}
    \bar{k}_3L\approx\pi n+i\ln\bigg|\frac{\e_\perp^{1/4}+\e_\perp^{-1/4}}{\e_\perp^{1/4}-\e_\perp^{-1/4}}\bigg|,\quad n\in \mathbb{Z},
\end{equation}
on the unphysical sheet. As for the physical sheet, the solutions obeying $|k_\perp/k_0|\ll1$ do not exist. The equations \eqref{res1} describe the resonances ($n=-\overline{\infty,-1}$) and the virtual state ($n=0$). At these points $\re k_0>0$ and $\im k_0\leqslant0$. The virtual state does not comply with the condition $|k_\perp/k_0|\ll1$ for $\re k_0>0$ and so it should be excluded. In order to observe the resonances in an experiment as a sharp maximum of the modulus of the transmission amplitude of the electromagnetic wave, it is necessary that $|\im k^n_0/\re k^n_0|\ll1$ and $|\im k^n_0/\re (k^n_0 -k^{n+1}_0)|\ll1$ (see, e.g., \cite{LandLifshQM.11}). The first requirement is always satisfied for sufficiently large $|n|$. For large $|n|$, we have
\begin{equation}
    \im k^n_0\approx-\frac{1}{\e_\perp L}\ln\bigg|\frac{\e_\perp^{1/4}+\e_\perp^{-1/4}}{\e_\perp^{1/4}-\e_\perp^{-1/4}}\bigg|.
\end{equation}
The second condition mentioned above leads to the restriction
\begin{equation}
    \frac{1}{\pi}\ln\bigg|\frac{\e_\perp^{1/4}+\e_\perp^{-1/4}}{\e_\perp^{1/4}-\e_\perp^{-1/4}}\bigg|\ll1.
\end{equation}
This condition is fulfilled when $\e_\perp\gtrsim5$ or $\e_\perp\lesssim1/5$. The more accurate estimate of the peak position of the absolute value of the transmission amplitude is obtained if one takes $|k_0^n|$ instead of  $\re k^0_n$ \cite{KlaiMois10}. Of course, in order to observe the resonance, it is necessary that its lifetime $\tau_l=1/|2\im k_0^n|$ be larger than the time of absorption of the photon in the medium $\tau_{abs}=|\e/(k_0\im\e)|$.

Apart from the paraxial approximation considered above, the cases when $|k_3/\bar{k}_3|\ll1$ for $\e_\perp>1$ or $|\bar{k}_3/k_3|\ll1$ for $\e_\perp<1$ are of interest (see Fig. \ref{Resons_plots}). In the first case, we find approximately
\begin{equation}
    \bar{k}_3L\approx\pi n -\frac{2ik^0_3}{\pi n}, \qquad k^0_3=\big[(\e_\perp^{-1}-1)k_\perp^2+\e_\perp^{-1}(\pi n/L)^2\big]^{1/2},
\end{equation}
on the unphysical sheet, where $n=\overline{-\infty,-1}$ and $k_3^0$ is assumed to be real. On the physical sheet, the bound states,
\begin{equation}
    \bar{k}_3L\approx\pi n +\frac{2k^0_3L}{\pi n},
\end{equation}
are present, where $k_3^0$ is real-valued and $n=\overline{1,\infty}$. In the second case, only the resonances on the unphysical sheet exist:
\begin{equation}
    \bar{k}_3L\approx\pi n -\frac{2\pi in}{k^0_3L},
\end{equation}
where $n=\overline{-\infty,-1}$. As we see, the resonances are close to the points $\bar{k}_3L\approx\pi |n|$ what is expected since the plate of a uniform isotropic dielectric is ideally transparent for those values of the energy.

The analysis of the approximate solutions to the second equation in \eqref{res_chol} is more intricate. We restrict ourselves to the case $|k_\perp/k_0|\ll1$. In this case, there are resonances on the unphysical sheet at the points
\begin{equation}\label{extraordinary_resonance}
    p_3L\approx\pi n+i\ln\bigg|\frac{\e_\parallel^{1/4}+\e_\parallel^{-1/4}}{\e_\parallel^{1/4}-\e_\parallel^{-1/4}}\bigg|,
\end{equation}
where $n=\overline{-\infty,-1}$. Supposing that $\re k_0>0$, $\im k_0<0$, and $|\im k_0/\re k_0|\ll1$, the approximate solution \eqref{extraordinary_resonance} can be cast into the form
\begin{equation}
    k_0=k'_{0}-ik''_{0},\qquad k'_{0}>0,\quad k''_{0}>0,
\end{equation}
where the real part of the energy satisfies the equation
\begin{equation}
    p_{3}\big|_{k_0=k'_{0}+i0}=\pi|n|/L,
\end{equation}
and the imaginary part is
\begin{equation}
    k''_{0}=\frac{1}{\tau_f L}\ln\bigg|\frac{\e_\parallel^{1/4}+\e_\parallel^{-1/4}}{\e_\parallel^{1/4}-\e_\parallel^{-1/4}}\bigg|.
\end{equation}
Here the notation has been introduced
\begin{equation}
    \tau_f:=\frac{\partial p_3}{\partial k_0}=\frac{q}{\pi}\int_{-\pi/(2q)}^{\pi/(2q)}\frac{dz}{\ups^{(2)}_3(z)},\qquad [\ups_3^{(2)}(z)]^{-1} :=\frac{\partial k_3^{(2)}(z)}{\partial k_0}.
\end{equation}
The dimensionless quantity $\tau_f$ equals the time of flight of the single cholesteric period by the extraordinary wave divided by the time of flight of the same distance by the photon in a vacuum. As wee see, the greater the time $\tau_f$, the stronger are the resonances. The resonances \eqref{res_chol} are observed in the intensity of radiation of a charged particle traversing the CLC plate as a wavy structure on the plot (see the figures in \cite{parax,wkb_chol}). For example, on the upper left plot in Fig. 2 of \cite{wkb_chol} the enlarged version of which is given on the right plot in Fig. \ref{Phi_dP_plots}, the period of oscillations of the radiation intensity is about $3.47\times 10^{-3}$ eV. This is in a good agreement with the period $3.49\times 10^{-3}$ eV following from Eq. \eqref{res1} for the parameters taken in Fig. 2 of \cite{wkb_chol}.

If the solutions entering into the matrix \eqref{alpha_k} were exact, condition \eqref{resonanses} would reproduce the bound states and the band spectrum found in the previous section. Namely, the boundaries of the bands are determined approximately by formula \eqref{bands}. When $n_\perp>1$ the modes stay in the CLC plate due to total internal reflection. For these modes $p_3L\approx\pi k$, $k=\overline{1,\infty}$, and the energy is given by formula \eqref{disper_law}. As a result, every band with the number $n$ turns into the set of discrete levels numerated by the index $k$. When $n_\perp<1$ the states found in the previous section are not bound in the CLC plate of a finite width, and the bands \eqref{bands} become a series of resonances. However, since we use the approximate mode functions in \eqref{alpha_k} without the subbarrier contributions described in Sec. \ref{Bound_Spec}, such a picture is not reproduced. In fact, the above analysis of resonances reveals only those resonances that are related to the total internal reflection of ordinary and extraordinary waves on the boundaries of the CLC plate.

\section{Conclusion}

Let us sum up the results. We investigated the solutions of the Maxwell equations in the CLC plate in the shortwave approximation. By using this approximation, we described the bound states \eqref{wkb_spectrum}, the resonances \eqref{res_chol}, the band structure \eqref{disper_law}, and the dispersion relations for photons \eqref{disper_law_bound} in such a plate. We showed that the photons moving almost perpendicular to the CLC axis are confined in the CLC plate. Their dispersion law along the CLC axis possesses approximately the form of the dispersion relation for a non-relativistic particle with a large mass \eqref{mass_eff} which is negative or positive depending on the band number. These modes resemble the bound states in a stack of thin-film waveguides. They can be excited by charged particles moving parallel and close to the surface of the CLC plate. The photons produced in such modes escape to a vacuum at the lateral faces of the CLC plate. They also can leave the CLC plate when its facets are tilted so as to avoid total internal reflection. Other methods for exciting and detection of such modes can be found in \cite{ZolKisSych}. 

The bound states we described produce sharp peaks in the density of photon states (see Fig. \ref{DOS_plots}) and so the quantum processes with photons evolving in the CLC plate are enhanced near these peaks. In the particular case of the plasma permittivity \eqref{permit_UV}, we found the dispersion laws of photons near the peaks of the density of states (see Fig. \ref{Disp_Rels_plots}). It turns out that the dispersion law of photons in the bound states in the direction perpendicular to the CLC axis takes the form of the dispersion relation for a massive relativistic particle \eqref{wkb-spectrum_plam} at sufficiently large photon energies. As for the small photon energies, it becomes linear \eqref{lin_disp_law} or quadratic \eqref{quadr_disp_law} depending on the sign of the anisotropy \eqref{anisotropy} of the CLC permittivity tensor.

As for resonances, we found simple equations \eqref{res_chol} for their positions and investigated the solutions of these equations in certain approximations (see Fig. \ref{Resons_plots}). These resonances reveal themselves, for example, in the intensity of transition radiation as a set of periodically arranged peaks \cite{parax,wkb_chol}. The theory developed in the present paper describes the period of these peaks rather accurately.

\paragraph{Acknowledgments.}

The reported study was supported by the Russian Ministry of Education and Science, the contract N 0721-2020-0033.

\appendix
\section{Unitarity relations}\label{Unit_Rels}

For the reader convenience, we will give in this appendix some basic formulas for the scattering problem on the line for the matrix Schr\"{o}dinger equation. Many of these formulas were presented in \cite{parax} but will also be given here as they are used in the main text. Furthermore, we will bring them into the form suitable for investigation of resonances.

\subsection{Complex-valued approach}

Let us consider the matrix differential equation of the second order
\begin{equation}\label{Schr_eqn}
    \big[\partial_zK(k_0;z)\partial_z+V(k_0;z)\big]u=0,
\end{equation}
where $K(k_0;z)$ and $V(k_0;z)$ are Hermitian $n\times n$ matrices for $k_0\in \mathbb{R}$, their elements being piecewise continuous functions. Suppose that
\begin{enumerate}
  \item $K(k_0;z)$ is nondegenerate and $K(k_0;z)|_{|z|\rightarrow\infty}$ is positive definite;
  \item $K(k_0;z)$ and $V(k_0;z)$ tend to constant values at $|z|\rightarrow\infty$ faster than $\exp(-c|z|)$ for any $c$ \cite{LandLifshQM.11};
  \item The eigenvalue problems,
\begin{equation}\label{eigen_prob}
    \big[-K(k_{0};z)\la^\pm_s(k_0)+V(k_{0};z)\big]_{z\rightarrow\pm\infty}f^\pm(s,k_0)=0,
\end{equation}
    where $s=\overline{1,n}$ numerates the solutions, possess the property that $\im\la^\pm_s(k_0)\neq0$ implies $\im k_0\neq0$.
\end{enumerate}
As for the Maxwell equations \eqref{Max_eqns2} describing the electromagnetic fields in a CLC plate of a finite width, these properties are met and $\la^\pm_s(k_0)=k_0^2-k_\perp^2$.

If $\psi$ and $\vf$ are solutions to \eqref{Schr_eqn}, then their complex Wronskian does not depend on $z$:
\begin{equation}\label{Wronsk_const}
    W[\vf,\psi]:=\vf^\dag K\partial_z\psi-\partial_z\vf^\dag K\psi=const.
\end{equation}
As long as
\begin{equation}
    W[\vf,\psi]=-W^*[\psi,\vf],
\end{equation}
the Wronskian defines an anti-Hermitian form on the solutions to Eq. \eqref{Schr_eqn}.

By the standard way, introduce the Jost functions
\begin{equation}\label{Jost_funcs}
    \big[\partial_zK(k_0;z)\partial_z+V(k_0;z)\big]F^\pm(s,k;z)=0,\qquad s=\overline{1,n},
\end{equation}
where
\begin{equation}\label{Jost_asympt}
    F^\pm(s,k;z)\underset{z\rightarrow\pm\infty}{\rightarrow}e^{ikz}f^\pm(s,k^2),\qquad \im k\geqslant0,
\end{equation}
and the vectors $f^\pm(s,k^2)$ satisfy Eqs. \eqref{eigen_prob} with
\begin{equation}
    k^2=\la^\pm_s(k_0).
\end{equation}
For real $k_0$, one can always put
\begin{equation}
    (f^\pm)^\dag(s,k_0)K\Big|_{z\rightarrow\pm\infty}f^\pm(s',k_0)=\de_{ss'}.
\end{equation}
Then it follows from the property \eqref{Wronsk_const} that for real $k$ and $k_0$,
\begin{equation}\label{scalar_prod}
    W[F^\pm(s,k),F^\pm(s',k)]=2ik\de_{ss'},\qquad W[F^\pm(s,k),F^\pm(s',-k)]=0,
\end{equation}
where for brevity we omit the argument $z$ of the Jost functions.

In virtue of the property 2 and the relations \eqref{scalar_prod}, the functions $F^+(s,\pm k)$ and $F^-(s,\pm k)$ constitute orthogonal bases in the $2n$-dimensional pseudo-Hilbert space of solutions of Eq. \eqref{Schr_eqn}. The two bases of solutions \eqref{Schr_eqn} are related as
\begin{equation}\label{2bases}
\begin{split}
    F^+(s,k)&=\al_{ss'}(k)F^-(s',k)+\be_{ss'}(k)F^-(s',-k),\\
    F^+(s,-k)&=\be_{ss'}(-k)F^-(s',k)+\al_{ss'}(-k)F^-(s',-k),
\end{split}
\end{equation}
where the summation over repeated indices is understood. Formulas \eqref{scalar_prod}, \eqref{2bases} imply that
\begin{equation}\label{unitar1}
    \al(k)\al^\dag(k)-\be(k)\be^\dag(k)=1,\qquad \al(-k)\be^\dag(k)-\be(-k)\al^\dag(k)=0,
\end{equation}
i.e.,
\begin{equation}
    \left[
      \begin{array}{cc}
        \al(k) & \be(k) \\
        \be(-k) & \al(-k) \\
      \end{array}
    \right]\in U(n,n).
\end{equation}

The relations \eqref{unitar1} are equivalent to unitarity of the corresponding scattering matrix. Namely, define the basis of solutions to Eq. \eqref{Schr_eqn} of the form
\begin{equation}\label{trans_refl}
\begin{split}
    \psi_1(s,k)&=t^{(1)}_{ss'}(k)F^+(s',k)=F^-(s,k)+r^{(1)}_{ss'}(k)F^-(s',-k),\\
    \psi_2(s,k)&=t^{(2)}_{ss'}(k)F^-(s',-k)=F^+(s,-k)+r^{(2)}_{ss'}(k)F^+(s',k).
\end{split}
\end{equation}
Then
\begin{equation}
\begin{gathered}
    t_{(1)}(k)=\al^{-1}(k),\qquad r_{(1)}(k)=\al^{-1}(k)\be(k),\qquad r_{(2)}(k)=-\be(-k)\al^{-1}(k),\\
    t_{(2)}^\dag(-k)=t_{(1)}(k),\qquad r_{(1,2)}^\dag(-k)=r_{(1,2)}(k),
\end{gathered}
\end{equation}
and the matrix
\begin{equation}
    S(k):=
    \left[
      \begin{array}{cc}
        t_{(1)}(k) & r_{(1)}(k) \\
        r_{(2)}(k) & t_{(2)}(k) \\
      \end{array}
    \right]
\end{equation}
is unitary.

The definition of the Jost functions and the relations \eqref{2bases} are valid for complex $k$ as well. It follows from \eqref{Jost_asympt}, \eqref{2bases} that the spectrum of the bound states is determined by the equation
\begin{equation}\label{resonanses}
    \det\al(k)=0,\qquad\im k>0,
\end{equation}
This condition ensures the existence of nontrivial linear combination of the Jost functions $F^+(s,k)$ that decreases exponentially to zero as $|z|\rightarrow\infty$. Assuming that Eq. \eqref{Schr_eqn} does not have square-integrable solutions for $k_0\not\in \mathbb{R}$, the property 3 implies that the spectrum of bound states lies only at $k=i|k|$. As for the Maxwell equations in a transparent medium, this assumption follows from the causality requirement (see, e.g., \cite{AbrGorDzyal,LandLifshST2,parax}).

The condition \eqref{resonanses} is useful for analysis of the spectrum of bound states. However, the coefficients $\al(k)$ and $\be(k)$ introduced above are not analytic functions of $k$, in general. This is because these coefficients are expressed in terms of the Jost functions, which are analytic functions of  $k$, by means of the complex Wronskian \eqref{Wronsk_const} that involves a complex conjugation. In order to obtain the analytic matrices $\al(k)$ and $\be(k)$ and to bring the formulas into a form completely analogous to the scattering theory for a scalar Schr\"{o}dinger equation on a line (see, e.g., \cite{Faddeev,FaddZakh,ZMNP,OstrElan05}), it is necessary to convert Eq. \eqref{Schr_eqn} into a real-valued form.

\subsection{Real-valued approach}

Taking the real and imaginary parts of  $K$, $V$, and $u$ in \eqref{Schr_eqn}, we arrive at the equation of the form \eqref{Schr_eqn} with orthogonal $2n\times 2n$ matrices $K_r$, $V_r$ and the column $u_r$ of $2n$ real functions. We also assume that the properties 1-3 are satisfied.

As regards the analytical structure of $K_r(k_0;z)$ and $V_r(k_0;z)$ for $|z|\rightarrow\infty$ and, consequently, of the eigenvalues $\la^\pm_{s}(k_0)$ and eigenfunctions $f^\pm_r(s,k^2)$, it can easily be investigated for a concrete scattering problem since the equations are greatly simplified at $|z|\rightarrow\infty$. In the case of the Maxwell equations \eqref{Max_eqns2}, the matrix $K_r(k_0,z)$ does not depend on $k_0$ for $|z|\rightarrow\infty$ and
\begin{equation}
    V_r(k_0;z)\underset{|z|\rightarrow\infty}{\rightarrow}k_0^2\tilde{V}_r,
\end{equation}
where $\tilde{V}_r$ is independent of $k_0$. Therefore, $\la^+_{s}(k_0)=\la^-_{s}(k_0)=k_0^2\tilde{\la}_{s}$, where $\tilde{\la}_{s}$ do not depend on $k_0$, and $f^\pm_r(s,k^2)=\tilde{f}^\pm_r(s)$ do not depend on $k^2$. Furthermore, for the problem at hand, the elements of the initial Hermitian matrices $K$ and $V$ are real at $|z|\rightarrow\infty$. Hence, the solutions of the eigenvalue problem \eqref{eigen_prob} for the matrices $K_r$ and $V_r$ can be written as
\begin{equation}\label{eigen_f}
    \tilde{f}^\pm_r(s)=[\tilde{f}^\pm(s),\tilde{f}^\pm(s)]/\sqrt{2},
\end{equation}
where $\tilde{f}^\pm_r(s)$ are the real-valued solutions to the problem \eqref{eigen_prob} with the matrices $K$ and $V$. Below, we suppose that all these relations are fulfilled.

The real Wronskian of two solutions of the matrix Sch\"{o}dinger equation does not depend on $z$,
\begin{equation}
    w[\vf,\psi]:=\vf^T K_r\partial_z\psi-\partial_z\vf^T K\psi=const.
\end{equation}
It defines the symplectic structure on the solutions to this equation. Introduce the Jost functions as in \eqref{Jost_funcs}, \eqref{Jost_asympt}, where
\begin{equation}
    k^2=k_0^2\tilde{\la}_s,
\end{equation}
and
\begin{equation}
    (\tilde{f}^\pm_r)^T(s)K_r\Big|_{z\rightarrow\pm\infty}\tilde{f}_r^\pm(s')=\de_{ss'}.
\end{equation}
In view of the properties 1, 2 and the obvious analyticity of $f^\pm_r(s,k^2)=\tilde{f}^\pm_r(s)$, the Jost functions are analytic functions of $k$ for $\im k\geqslant0$, except for the point $k=0$ \cite{LandLifshQM.11}, and
\begin{equation}\label{symm_prop_1}
    [F^\pm_r(s,k;z)]^*=F^\pm_r(s,-k^*;z).
\end{equation}

The relations,
\begin{equation}\label{scalar_prod1}
    w[F^\pm_r(s,k),F^\pm_r(s',-k)]=-2ik\de_{ss'},\qquad w[F^\pm_r(s,k),F^\pm_r(s',k)]=0,
\end{equation}
hold on the physical sheet $\im k\geqslant0$. The property 2 and relations \eqref{scalar_prod1} imply that the functions $F^+_r(s,\pm k)$ and $F^-_r(s,\pm k)$ constitute symplectic bases in the $4n$-dimensional symplectic space of solutions to Eq. \eqref{Schr_eqn}. These bases are related as
\begin{equation}\label{2bases1}
\begin{split}
    F^+_r(s,k)&=\Phi_{ss'}(k)F^-_r(s',k)+\Psi_{ss'}(k)F^-_r(s',-k),\\
    F^+_r(s,-k)&=\Psi_{ss'}(-k)F^-_r(s',k)+\Phi_{ss'}(-k)F^-_r(s',-k).
\end{split}
\end{equation}
This transform is symplectic provided that
\begin{equation}
\begin{aligned}
    \Phi(k)\Phi^T(-k)-\Psi(k)\Psi^T(-k)&=1,&\qquad\Phi(k)\Psi^T(k)&=\Psi(k)\Phi^T(k),\\
    \Phi^T(k)\Phi(-k)-\Psi^T(-k)\Psi(k)&=1,&\qquad\Phi^T(k)\Psi(-k)&=\Psi^T(-k)\Phi(k).
\end{aligned}
\end{equation}
The above relations imply
\begin{equation}\label{sympl_rels}
\begin{aligned}
    \Phi^{-1}(k)\Psi(k)&=[\Phi^{-1}(k)\Psi(k)]^T,&\quad \Psi(-k)\Phi^{-1}(k)&=[\Psi(-k)\Phi^{-1}(k)]^T,\\
    [\Phi^T(-k)\Phi(k)]^{-1}&+\Phi^{-1}(k)\Psi(k)\Phi^{-1}(-k)\Psi(-k)=1,&\quad [\Phi(-k)\Phi^T(k)]^{-1}&+\Psi(-k)\Phi^{-1}(k)\Psi(k)\Phi^{-1}(-k)=1.
\end{aligned}
\end{equation}
Using the skew-symmetric products \eqref{scalar_prod1}, the matrices $\Phi(k)$ and $\Psi(k)$ are expressed through the Jost functions from Eq. \eqref{2bases1}. Then it follows from these expressions and the property \eqref{symm_prop_1} that the matrices $\Phi(k)$ and $\Psi(k)$ are analytic functions for $\im k\geqslant0$ with the exception of the point $k=0$ and
\begin{equation}\label{symm_prop_2}
    \Phi^*(k)=\Phi(-k^*),\qquad \Psi^*(k)=\Psi(-k^*).
\end{equation}

Introducing the transmission $T_{(1,2)}(k)$ and reflection $R_{(1,2)}(k)$ matrices as in \eqref{trans_refl} and employing \eqref{sympl_rels}, we deduce
\begin{equation}
    T_{(1)}(k)=\Phi^{-1}(k),\qquad T_{(2)}(k)=[\Phi^{-1}(k)]^T,\qquad R_{(1)}(k)=\Phi^{-1}(k)\Psi(k),\qquad R_{(2)}(k)=-\Psi(-k)\Phi^{-1}(k).
\end{equation}
These matrices obey the relations
\begin{equation}
    R_{(1,2)}^T(k)=R_{(1,2)}(k),\qquad T_{(1)}^T(k)=T_{(2)}(k),
\end{equation}
the symmetry property of the form \eqref{symm_prop_2}, and the unitarity relations for the scattering matrix
\begin{equation}
    S_r(k):=
    \left[
      \begin{array}{cc}
        T_{(1)}(k) & R_{(1)}(k) \\
        R_{(2)}(k) & T_{(2)}(k) \\
      \end{array}
    \right],
\end{equation}
for real $k$.

The interpretation of zeros of the quantity $\det\Phi(k)$ on the physical sheet is the same as for $\det\al(k)$. Moreover, the zeros of $\det\Phi(k)$ on the unphysical sheet, $\im k<0$, define resonances and virtual states. The virtual (anti-bound) states have $\re k=0$. The branch points of $\det\Phi(k)$ for $k\geqslant0$ correspond to the boundaries of the continuous spectrum. Thus, the spectrum of the bound states is found from the condition
\begin{equation}\label{resonanses1}
    \det\Phi(k)=0,\qquad\im k>0,
\end{equation}
and is located at $k=i|k|$. The change of the density of states due to nontrivial scattering has the form \cite{Dew_scat,Bordag_scat}
\begin{equation}\label{dens_of_states}
    \De\rho(\la)=-\frac{i}{2\pi}\partial_\la\ln\det S_r(\sqrt{\la})=\frac{i}{2\pi}\ln\frac{\det\Phi(\sqrt{\la_+})}{\det\Phi(\sqrt{\la_-})},
\end{equation}
where $\la_\pm=\la\pm i0$, $\la\in \mathbb{R}$, and the branch of the square root is taken with the cut along the real positive semiaxis. In particular, it is seen from this expression that if the resonance is close to the point $\la'>0$ on the real axis, then there is an increase of the density of states in the vicinity of $\la'$.

Now we show that under the assumptions mentioned above
\begin{equation}\label{a_Ph_b_Ps}
    \al(k)=\Phi(k),\qquad \be(k)=\Psi(k).
\end{equation}
Indeed, let us write the Jost functions in the block form
\begin{equation}
    F^\pm_r(s,k)=\big[\re F^\pm_1(s,k)+i \im F^\pm_1(s,k),\re F^\pm_2(s,k)+i \im F^\pm_2(s,k)\big].
\end{equation}
It follows from \eqref{eigen_f} that
\begin{equation}
    \re F^\pm_{1,2}(s,k)+i \im F^\pm_1(s,k)\underset{z\rightarrow\pm\infty}{\rightarrow} e^{ikz}\tilde{f}^\pm(s)/\sqrt{2}.
\end{equation}
It is also clear that by construction
\begin{equation}
    \re F^\pm_1(s,k)+i \re F^\pm_1(s,k),\qquad \im F^\pm_2(s,k)+i \im F^\pm_2(s,k)
\end{equation}
are the solutions to Eq. \eqref{Schr_eqn}. Then $F^\pm_1(s,k)+i F^\pm_2(s,k)$ is also the solution to Eq. \eqref{Schr_eqn} and
\begin{equation}
    F^\pm_1(s,k)+i F^\pm_2(s,k) \underset{z\rightarrow\pm\infty}{\rightarrow} \frac{1+i}{\sqrt{2}}e^{ikz}\tilde{f}^\pm(s).
\end{equation}
Therefore,
\begin{equation}
    F^\pm_1(s,k)+i F^\pm_2(s,k)=e^{i\pi/4} F^\pm(s,k).
\end{equation}
Using this relation, we see that Eqs. \eqref{2bases1} turn into \eqref{2bases} and the equalities \eqref{a_Ph_b_Ps} are satisfied.

\section{Joining the mode functions}\label{Join_Coeff_App}

The below results are borrowed from \cite{wkb_chol} and are given here for the reader convenience. It is useful to rewrite the equations for the coefficients of the linear combination of solutions \eqref{mode_func_cho}, \eqref{mode_funcz<mL} of the Maxwell equations that arise on imposing the boundary conditions \eqref{bound_conds_chol} in the form of the matrix equation
\begin{equation}\label{joining_eqn}
    \left[
       \begin{array}{cc}
         U & 0 \\
         UT & -HT' \\
       \end{array}
     \right]
     \left[
       \begin{array}{c}
         a_{ch} \\
         a_l \\
       \end{array}
     \right]=
     \left[
       \begin{array}{c}
         g \\
         0 \\
       \end{array}
     \right],
\end{equation}
where
\begin{equation}
    a^T_{ch}=(r_1,r_2,l_1,l_2),\qquad a_l^T=(d_+,d_-,h_+,h_-),
\end{equation}
and
\begin{equation}\label{joining_eqn_3}
\begin{gathered}
    T=\diag(e^{-i\bar{k}_3L},e^{-ip_3L},e^{i\bar{k}_3L},e^{ip_3L}),\qquad    T'=\diag(e^{-ik_3L},e^{-ik_3L},e^{ik_3L},e^{ik_3L}),\\
    H=\frac{1}{\sqrt{2}}
    \left[
      \begin{array}{cccc}
        \cos\theta-1 & \cos\theta+1 & -\cos\theta-1 & -\cos\theta+1 \\
        \cos\theta+1 & \cos\theta-1 & -\cos\theta+1 & -\cos\theta-1 \\
        k_0(1-\cos\theta) & k_0(1+\cos\theta) & k_0(1+\cos\theta) & k_0(1-\cos\theta) \\
        k_0(1+\cos\theta) & k_0(1-\cos\theta) & k_0(1-\cos\theta) & k_0(1+\cos\theta) \\
      \end{array}
    \right],\\
    U=
    \left[
      \begin{array}{cc}
        U_{11} & U_{12} \\
        U_{21} & U_{22} \\
      \end{array}
    \right],\qquad
    U_{11}=
    \left[
      \begin{array}{cc}
        a_+^{(1)} & a_+^{(2)} \\
        a_-^{(1)} & a_-^{(2)} \\
      \end{array}
    \right],\qquad U_{12}=U_{11}\Big|_{a\rightarrow b},\\
    U_{21}=-iK \partial_zU_{11},\qquad U_{22}=-iK \partial_zU_{12}=-U_{21}\Big|_{a\rightarrow b},\\
    g^T=\big[\cos\theta-s,\cos\theta+s,k_0(1-s\cos\theta),k_0(1+s\cos\theta)\big]/\sqrt{2},
\end{gathered}
\end{equation}
where $a_\pm^{(1,2)}$, $b_\pm^{(1,2)}$ are the solutions \eqref{ordinary_+}-\eqref{extraordinary_-}. All the functions of $z$ entering into the components of the matrix $U$ are taken at $z=0$. Explicitly,
\begin{equation}
    U_{21}=
    \left[
      \begin{array}{cc}
        \bar{k}_3\big[a_+^{(1)}+\frac{k_\perp^2}{2\bar{k}_3^2}(a_+^{(1)}+a_-^{(1)})\big] & k^{(2)}_3\big[a_+^{(2)}+\frac{k_\perp^2}{2\bar{k}_3^2}(a_+^{(2)}+a_-^{(2)})\big] \\
        \bar{k}_3\big[a_-^{(1)}+\frac{k_\perp^2}{2\bar{k}_3^2}(a_-^{(1)}+a_+^{(1)})\big] & k^{(2)}_3\big[a_-^{(2)}+\frac{k_\perp^2}{2\bar{k}_3^2}(a_-^{(2)}+a_+^{(2)})\big] \\
      \end{array}
    \right].
\end{equation}
Then we come to
\begin{equation}\label{joining_sol}
    a_{ch}=U^{-1}g,\qquad a_l=(T')^{-1}H^{-1}UTU^{-1}g,
\end{equation}
for the solution of Eq. \eqref{joining_eqn}.

Despite the fact that the solutions \eqref{ordinary_+}-\eqref{extraordinary_-} are approximate solutions of the Maxwell equations, the unitarity relation \eqref{unitar1} for the coefficients \eqref{joining_sol} holds exactly:
\begin{equation}
    1+|h_+|^2+|h_-|^2=|d_+|^2+|d_-|^2.
\end{equation}
The resulting expression for the mode function $\bs{\psi}(s,\spk;x)$ must be normalized. The normalization condition leads to (see for details, Sec. 5.A of \cite{BKL5})
\begin{equation}\label{norm_const0}
    |c|^2=2(1+a^\dag_l a_l)^{-1}=\big(|d_+|^2+|d_-|^2\big)^{-1}.
\end{equation}
It is not difficult to show \cite{wkb_chol} that the variable $\vf$ appears in $|c|^2$ only as the combination $s\vf$.

The expressions for the coefficients $a_{ch}$ of the linear combination become
\begin{equation}\label{ach_wkb}
\begin{split}
    r_1&=is\sqrt{\frac{k_0}{\e_\perp-n_\perp^2\cos^2\vf}}(\bar{n}_3+\cos\theta)\frac{\bar{n}_3(z_s^2+1) -(z_s^2-1)(\bar{n}_3\cos\theta+\sin^2\theta)}{4\bar{n}_3^{1/2}z_s},\\
    r_2&=\sqrt{\frac{k_0}{\e_\perp-n_\perp^2\cos^2\vf}}e^{-iS(-\vf)} \frac{(z_s^2+1)(\bar{n}_3^2+\e_\perp n_3^{(2)}\cos\theta)-\e_\perp(z_s^2-1)(n_3^{(2)}+\cos\theta)}{4\e_\perp^{1/2}(n_3^{(2)})^{1/2} z_s},\\
    l_1&=is\sqrt{\frac{k_0}{\e_\perp-n_\perp^2\cos^2\vf}}(\bar{n}_3-\cos\theta)\frac{\bar{n}_3(z_s^2+1) -(z_s^2-1)(\bar{n}_3\cos\theta-\sin^2\theta)}{4\bar{n}_3^{1/2}z_s},\\
    l_2&=-\sqrt{\frac{k_0}{\e_\perp-n_\perp^2\cos^2\vf}}e^{iS(-\vf)} \frac{(z_s^2+1)(\bar{n}_3^2-\e_\perp n_3^{(2)}\cos\theta)+\e_\perp(z_s^2-1)(n_3^{(2)}-\cos\theta)}{4\e_\perp^{1/2}(n_3^{(2)})^{1/2} z_s},
\end{split}
\end{equation}
where $\bar{n}_3:=\bar{k}_3/k_0$ and $n^{(2)}_3:=k^{(2)}_3/k_0$. The normalization constant reads
\begin{equation}\label{norm_const}
    |c|^{-2}=\nu_0+\big[\nu_1 e^{-i(\bar{k}_3+p_3)L} +\nu_2 e^{-2i\bar{k}_3L} +\nu_3 e^{-2ip_3L} +\nu_4 e^{i(\bar{k}_3-p_3)L} +c.c.\big],
\end{equation}
where
\begin{fleqn}
\begin{equation}
\begin{split}
    \nu_0=\,&\big[8\bar{n}^2_3(n_3^{(2)})^2(1-n_\perp^2)\e_\perp^2(4\e_\perp z_s^2-n_\perp^2(z_s^2+1)^2)^2 \big]^{-1} \bigg\{\e_\perp^4(n_3^{(2)})^4\bar{n}_3^2[4z^2_s-n_\perp^2(z_s^2+1)^2]^2+\\
    &+\bar{n}_3^2\big[\e_\perp^2(4z^2_s+n_\perp^2(z^2_s-1)^2) -2\e_\perp n_\perp^2(z^2_s+1)+n_\perp^4(z_s^4+1)^2\big]^2+\\ &+\e_\perp^2(n_3^{(2)})^2\Big[ \e_\perp^4\big(4z_s^2+n_\perp^2(z_s^2-1)^2\big)^2 -8\e_\perp^3\big[n_\perp^4(z_s^4-1)^2+2n_\perp^2(z_s^6+14z_s^4+z_s^2) -24z_s^4\big]+\\
    &+\e_\perp^2\big[6n_\perp^6(z_s^4-1)^2+n_\perp^4(13z_s^8+124z_s^6+414z_s^4+124z_s^2+13) -8n_\perp^2z_s^2(13z_s^4+50z_s^2+13)+16z_s^4\big]-\\
    &-2\e_\perp n_\perp^2\big[n_\perp^4(z_s^2+1)^4(13z_s^4+58z_s^2+13) -4n_\perp^2(z_s^2+1)^2(z_s^4+18z_s^2+1) -16z_s^4 \big]+\\ &+14n_\perp^8(z_s^2+1)^4 -8n_\perp^6(z_s^2+1)^2(z_s^4+5z_s^2+1) +16n_\perp^4z_s^4 \Big] \bigg\},
\end{split}
\end{equation}
\begin{equation}
\begin{split}
    \nu_1=\,&\frac{\chi_\perp n_\perp^4(z_s^4-1)(\bar{n}_3+\e_\perp n_3^{(2)})}{8\bar{n}_3n^{(2)}_3(1-n_\perp^2)\e_\perp(4z_s^2\e_\perp -n_\perp^2(z_s^2+1)^2)^2}\big[\e_\perp^2(1-n_\perp^2)(z_s^2-1)^2 -\bar{n}_3^4(z_s^2+1)^2+\\
    &+\chi_\perp n_\perp^2(z_s^4-1)(\bar{n}_3+\e_\perp n_3^{(2)}) +\bar{n}_3n_3^{(2)}\e_\perp(4z_s^2-n_\perp^2(z_s^2+1)^2) \big],
\end{split}
\end{equation}
\begin{equation}
    \nu_2=-\frac{\chi_\perp^2 \big[4z_s^2\bar{n}_3^2+\e_\perp n_\perp^2(z_s^2-1)^2-2\bar{n}_3n_\perp^2(z_s^4-1) \big]\big[4z_s^2\bar{n}_3^2+\e_\perp n_\perp^2(z_s^2-1)^2 \big]}{16\bar{n}_3^2(1-n_\perp^2)(4z_s^2\e_\perp -n_\perp^2(z_s^2+1)^2)^2},
\end{equation}
\begin{equation}
\begin{split}
    \nu_3=\,&-\frac{\big[\bar{n}_3^4(z_s^4+1)^2-\e_\perp^2(n_3^{(2)})^2(4z_s^2-n_\perp^2(z_s^2+1)^2)-\e_\perp^2(1-n_\perp^2)(z_s^2-1)^2 \big]}{16(n^{(2)}_3)^2(1-n_\perp^2)\e_\perp^2(4z_s^2\e_\perp -n_\perp^2(z_s^2+1)^2)^2}\times\\
    &\times \big[\bar{n}_3^4(z_s^4+1)^2-\e_\perp^2(n_3^{(2)})^2(4z_s^2-n_\perp^2(z_s^2+1)^2)-\e_\perp^2(1-n_\perp^2)(z_s^2-1)^2 -2n_3^{(2)}n_\perp^2\chi_\perp\e_\perp(z_s^4-1) \big],
\end{split}
\end{equation}
\begin{equation}
\begin{split}
    \nu_4=\,&\frac{\chi_\perp n_\perp^2(\bar{n}_3-\e_\perp n_3^{(2)})(z_s^4-1)}{8\bar{n}_3n_3^{(2)}(1-n_\perp^2)\e_\perp(4z_s^2\e_\perp-n_\perp^2(z_s^2+1)^2)^2}
    \big[\bar{n}_3 (z_s^2+1)^2(\e_\perp n_\perp^2 n_3^{(2)}-\bar{n}_3^3)-\\
    &-\chi_\perp n_\perp^2(z_s^4-1) (\bar{n}_3-\e_\perp n_3^{(2)}) -4\bar{n}_3n_3^{(2)}\e_\perp z_s^2 +(1-n_\perp^2)(z_s^2-1)^2\e_\perp^2\big],
\end{split}
\end{equation}
\end{fleqn}
where $z_s:=e^{is\vf}$ and $\chi_{\perp,\parallel}:=\e_{\perp,\parallel}-1$.



\end{document}